\documentclass[a4paper,11pt]{article}
\pdfoutput=1 

\usepackage{jheppub} 

\usepackage{graphics}
\usepackage{amssymb}
\usepackage{amsmath}
\usepackage{bbold}
\usepackage{hyperref}
\usepackage{slashed}
\usepackage[normalem]{ulem}
\usepackage{orcidlink}

\allowdisplaybreaks

\newcommand{\ord}{\mathcal{O}}

\newcommand{\bnslash}{\bar{n}\!\!\!\slash}
\newcommand{\nslash}{n\!\!\!\slash}
\newcommand{\cP}{{\mathcal P}}
\newcommand{\bn}{\bar{n}}
\newcommand{\bnP}{\overline {\mathcal P}}

\newcommand{\msb}{{\overline{\mathrm{MS}}}}
\newcommand{\eps}{\epsilon}
\newcommand{\nn}{\nonumber}

\DeclareRobustCommand{\eq}[1]{eq.~\eqref{eq:#1}}
\DeclareRobustCommand{\eqs}[2]{eqs.~\eqref{eq:#1} and \eqref{eq:#2}}
\DeclareRobustCommand{\eqsm}[2]{eqs.~\eqref{eq:#1}\,--\,\eqref{eq:#2}}

\DeclareRobustCommand{\fig}[1]{figure~\ref{fig:#1}}
\DeclareRobustCommand{\figs}[2]{figures~\ref{fig:#1} and \ref{fig:#2}}
\DeclareRobustCommand{\sec}[1]{section~\ref{sec:#1}}

\DeclareRobustCommand{\app}[1]{appendix~\ref{app:#1}}

\DeclareRobustCommand{\rcite}[1]{ref.\,\cite{#1}}
\DeclareRobustCommand{\rcites}[1]{refs.\,\cite{#1}}

\newcommand{\ri}{\mathrm{i}}

\newcommand{\rd}{\mathrm{d}}

\newcommand{\bare}{\mathrm{bare}}

\newcommand{\tr}{\mathrm{tr}}

\newcommand{\cB}{{\mathcal B}}
\newcommand{\cL}{{\mathcal L}}

\newcommand{\Mae}[3]{\bigl\langle#1\bigr\rvert#2\bigr\rvert#3\bigr\rangle}

\newcommand{\Li}{\mathrm{Li}}
\newcommand{\w}{\omega}

\newcommand{\convz}{\!\otimes_z\!}
\newcommand{\convx}{\!\otimes_x\!}

\newcommand{\zero}{{(0)}}
\newcommand{\one}{{(1)}}
\newcommand{\two}{{(2)}}

\newcommand{\oq}{{\mathcal{Q}}}

\newcommand{\lqcd}{\Lambda_\mathrm{QCD}}

\newcommand{\SCETI}{SCET$_\mathrm{I}$}
\newcommand{\SCETII}{SCET$_\mathrm{II}$}

\DeclareMathAlphabet{\mathbbold}{U}{bbold}{m}{n}

\newcommand{\sumintX}{\;\underset{\!\!\!X}{\int\!\!\!\!\!\!\!\!\!\sum}}

\title{
NNLO electron structure functions (PDFs) from SCET\!
}

\preprint{
\begin{flushright}
FR-PHENO-2025-009
\end{flushright}
}

\author[]{Maximilian Stahlhofen\orcidlink{0000-0002-2613-9014}}

\affiliation[]{Albert-Ludwigs-Universit\"at Freiburg, Physikalisches Institut, 79104 Freiburg, Germany}

\emailAdd{maximilian.stahlhofen@physik.uni-freiburg.de}

\abstract{
We calculate the electron structure functions, aka parton distribution functions (PDFs), to NNLO in QED.
The calculation is based on the definition of the PDFs in terms of operator matrix elements in soft collinear effective theory (SCET) and directly performed in momentum space.
The electron PDFs describe the universal effects of collinear initial state radiation (ISR) off the electron (or positron) in DIS or $e^+ e^-$ collision events.
The parton collinear to the electron that enters the hard scattering process governed by the energy scale $Q$ can be a photon or a (anti)fermion with  mass $m_f \sim m_e \ll Q$.
Our SCET momentum-space calculation confirms earlier results obtained from a QED computation in Mellin space and extends them by taking more than one massive fermion flavor into account.
The well-known factorization of sufficiently inclusive cross sections into PDFs with massive fermions and hard partonic cross sections with massless fermions  at leading order in $m_e/Q$ is discussed from the SCET perspective.
Analogies to the nonperturbative PDFs and collinear factorization in QCD are pointed out.
Based on this factorization DGLAP-type resummation of large logarithms of the ratio $m_e/Q$ is straightforward.
}

\begin{document}
\maketitle
\flushbottom

\section{Introduction}
\label{sec:intro}

Factorization plays a crucial role in the theoretical description of high-energy scattering processes.
A prominent example in this context is the factorization of inclusive hadronic cross sections, as measured e.g.\ in proton-proton collisions at the LHC: These cross sections take the form of a convolution of a partonic cross section with two parton distribution functions (PDFs) at leading order in $\lqcd/Q$, where $Q \gg \lqcd$ represents the energy scale of all kinematic invariants of the partonic process. The process-independent PDFs quantify the parton content of the hadrons as a function of the light-cone momentum fraction $x$ (Bjorken variable) carried by the parton w.r.t.\ the incoming hadron momentum.

Similarly, this kind of ``collinear factorization'' can also be established for lepton-hadron or lepton-lepton scattering processes.
The PDFs describing the parton content as well as the accompanying collinear initial-state radiation (ISR) of the incoming (unpolarized) leptons are for historical reasons often called lepton ``structure functions''~\cite{Kuraev:1985hb}.
Following more recent literature~\cite{Frixione:2023gmf,Garosi:2023bvq,Han:2021kes,Han:2020uid,Bertone:2022ktl,Bertone:2019hks,Frixione:2019lga} (and despite the title), we however prefer to use the term ``lepton PDF'' (or simply PDF if there is no danger of confusion)%
\footnote{When referring to the distribution of a specific parton $i$ (e.g.\ a lepton, photon, etc.) inside a lepton ($\ell$), we will use the term ``parton-in-lepton'' PDF and denote it as $f_{i/\ell}$.}
in this paper in order to emphasize the close analogy to the usual hadron PDFs.%
\footnote{See e.g.\ \rcite{Bertone:2019hks} for a critical account of the terminology.}
In contrast to their hadronic analogs the lepton PDFs are governed by the mass of the lepton (rather than $\lqcd$) as the characteristic energy scale and, most notably, can be calculated perturbatively from first principles in QED.%
\footnote{At higher orders in a full Standard Model calculation of course also nonperturbative corrections involving the hadronization scale $\lqcd \ll Q$ come into play.}
Just like for hadron PDFs, the partons of a lepton PDF can in principle be all (beyond-) Standard Model (anti-)particles as long as their masses are substantially smaller than the hard scale $Q$.
To next-to-next-to-leading order (NNLO) in the perturbative expansion and excluding heavy vector bosons, however, only photons, leptons and light quarks occur. The latter two are jointly referred to as fermions of different flavor ($f$) in the following.
Along the same lines, one can also define a photon PDF describing the parton content and collinear ISR of an incoming photon originating from a dedicated photon beam or the beamstrahlung of a lepton beam.
The characteristic energy scale of the photon PDF is set by the masses of the (light) fermions involved in the leading photon splitting amplitudes.

For the important case of $e^+ e^-$ collisions collinear factorization in QED concretely predicts that sufficiently inclusive cross sections take the form
\begin{align}
  \rd \sigma_{e\bar{e}} = \sum_{a,b} \int\!\! \rd x_a \int\!\! \rd x_b\;
  f_{a/e}(x_a,m,\mu) \; f_{b/\bar{e}}(x_b,m,\mu) \;
  \rd \hat{\sigma}_{ab} (x_a, x_b, Q, \mu)
  \;+\; \ord \biggl(\frac{m}{Q} \biggr)
  \,.
  \label{eq:fact}
\end{align}
Here and in the following $m$ refers to the scale of the light fermion masses ($m_f \sim m_e \sim m$) and $Q \gg m$ to the common energy scale assumed for all other dimensionful parameters governing the scattering process (like the center-of-mass energy, heavy particle masses, Born-level kinematic invariants, or phase space constraints by the measurement).%
\footnote{This does of course not exclude the possibility of a large internal hierarchy between the scales ($\gg m$) encompassed in our notation by Q. In that case $\hat{\sigma}_{ab}$  may  allow for further factorization.}
The electron and positron PDFs encode the effects of collinear ISR and are denoted by $f_{a/e}$ and $f_{b/\bar{e}}$, respectively, where the first indices ($a$,$b$) indicate the respective partons initiating the hard (short-distance) scattering process at the scale $Q$.
The ``partonic cross section'' $\hat{\sigma}_{ab}$ describes this hard scattering and is computed in the \textit{massless} limit, i.e.\ for $m_a=m_b=m_e = m_f =0$, and with on-shell external particles.
The sum in \eq{fact} is taken over all types of partons ($a$, $b$) in the electron and the positron, while the integrations are over the fractions ($x_a$, $x_b$) of the electron and positron momenta carried by the respective partons.
This kind of factorization is also referred to as the ``structure function approach''~\cite{Kuraev:1985hb} to account for QED ISR, see \rcites{Frixione:2022ofv,Denner:2019vbn} for reviews.

Note that the (schematic) factorization formula in \eq{fact} only holds at ``leading power'', i.e.\ at leading order of the expansion in $m/Q$ and if the partonic cross section $\hat{\sigma}_{ab}$ is infrared (IR) finite apart from the collinear divergences associated with the initial state partons $a$ and $b$ in the massless limit.
The latter divergences cancel with corresponding ultraviolet (UV) divergences of the IR finite PDFs.
The requirement that additional soft and collinear singularities, that may occur in individual Feynman diagrams contributing to $\hat{\sigma}_{ab}$, cancel due to the KLN theorem~\cite{Lee:1964is,Kinoshita:1962ur} is implied by the ``sufficient inclusiveness'' of the process as assumed above.
For more exclusive processes the factorization formula in \eq{fact} must be supplemented by additional collinear and/or soft factorization functions.
For example, when the energy $\sim Q$ of a final-state photon or light lepton is constrained such that an additional collinear (mass) singularity arises, the leading power cross section involves also a convolution with a corresponding fragmentation function, the final-state analog of the PDF~\cite{Collins:1981uw}.
For later reference we also note that a related factorization formula holds for massive (QED) amplitudes. It allows for the computation of the virtual corrections to a massive (i.e.\ $m_f$-dependent) amplitude (e.g.\ the electron Dirac form factor) at leading power in terms of the corresponding massless loop amplitude and a ``massification'' factor, see \rcites{Wang:2023qbf,Engel:2018fsb,Becher:2007cu,Mitov:2006xs,Catani:2000ef}.

In the present paper we calculate the electron PDFs $f_{i/e}$, where $i$ stands for an arbitrary (light) fermion ($i=f$), antifermion ($i=\bar{f}$), or photon ($i=\gamma$), at NNLO in QED with an arbitrary number of fermion flavors.
Due to charge conjugation symmetry of QED the corresponding positron PDFs are then given by $f_{i/\bar{e}} = f_{\bar{i}/e}$ (with $\overline{\gamma}\equiv \gamma$).
The calculation is performed directly in momentum (``$x$''-)space and based on the definition of the PDFs in terms of operator matrix elements in soft-collinear effective theory (SCET)~\cite{Bauer:2000ew, Bauer:2000yr,
Bauer:2001ct, Bauer:2001yt, Bauer:2002nz, Beneke:2002ph}.%
\footnote{We note that SCET also allows one to systematically compute the power corrections to factorization formulae like \eq{fact}, if ever needed.}
The NNLO results for the PDFs $f_{e/e}$, $f_{\bar{e}/e}$, and $f_{\gamma/e}$ in QED with only a single fermion flavor ($f=e$) were already obtained quite some time ago in \rcites{Blumlein:2011mi,Ablinger:2020qvo}, respectively.
In those works the (operator matrix elements of the) PDFs were defined and computed in Mellin (moment) space, where (Mellin) convolutions like in \eq{fact} turn into ordinary products. To arrive at the final expression for the cross section or the $x$-space PDFs one then has to perform an inverse Mellin transform. This has in fact been done for the PDFs in \rcites{Blumlein:2011mi,Ablinger:2020qvo}.
Curiously, due to a disagreement between \rcite{Blumlein:2011mi} with \rcite{Berends:1987ab} on the ISR corrections to the
(Drell--Yan type) process $e^+ e^- \to \gamma^*/Z^*$, \rcite{Blumlein:2011mi} raised some doubts on the validity of the factorization according to \eq{fact} (despite its conceptual foundations known from QCD~\cite{Collins:1989gx}).
In \rcite{Blumlein:2020jrf}, however, the results of \rcite{Blumlein:2011mi} were confirmed by a corresponding NNLO QED calculation with full electron mass dependence and the source of discrepancy could be traced back to missing/erroneous terms in \rcite{Berends:1987ab}.
In \sec{calculation} we recalculate $f_{e/e}$, $f_{\bar{e}/e}$ explicitly in momentum space and once again confirm the results of \rcite{Blumlein:2011mi}. For part of the purely virtual two-loop contribution we employ a known massification factor~\cite{Becher:2007cu}. The one-flavor QED ($f=e$) result for $f_{\gamma/e}$  we take from \rcite{Ablinger:2020qvo} and check the agreement with the SCET formulation via a sum rule relation in \sec{results}.
We also compute for the first time (besides $f_{f/e}$, $f_{\bar{f}/e}$ with $f\neq e$)  all NNLO corrections to $f_{e/e}$, $f_{\bar{e}/e}$, and $f_{\gamma/e}$ arising from adding additional fermion flavors.
In \rcite{Frixione:2019lga} the NLO expressions for the $x$-space PDFs have been extracted from a comparison of massless and massive QED cross sections (in a suitable collinear limit) by demanding factorization. The $\msb$ renormalized results agree with those given (much) earlier in \rcite{Blumlein:2011mi}.%
\footnote{It seems that the author of \rcite{Frixione:2019lga} was not aware of \rcite{Blumlein:2011mi}, because it is not quoted by \rcite{Frixione:2019lga} even though it contains not only the complete set of NLO PDFs, but also NNLO results for $f_{e/e}$ and $f_{\bar{e}/e}$.}

As stated above, the bare PDFs are UV divergent and the bare partonic cross section contains collinear IR divergences such that the physical cross section in \eq{fact} is finite. From the effective field theory (EFT) perspective these singularities are a consequence of the manifest separation of the scales $m$ and $Q$ and can be used to resum large logarithms of the ratio $m/Q$ by means of renormalization group (RG) evolution to all orders in the electromagnetic coupling constant $\alpha$.
Upon renormalization of the PDF operators and the partonic cross section, which in the EFT picture corresponds to a Wilson coefficient, in a suitable (subtraction) scheme like $\msb$,  $f_{i/j}$ and $\hat{\sigma}_{ab}$ depend on a renormalization scale $\mu$ as indicated in \eq{fact}.
Their $\mu$-evolution, i.e.\ their dependence on the unphysical parameter $\mu$, is determined by their respective renormalization group equation (RGE).
The RGEs of the QED PDFs correspond to the QED version of the famous DGLAP equations~\cite{Altarelli:1977zs,Dokshitzer:1977sg,Gribov:1972ri} in QCD and can be directly derived from the latter by simply adjusting the color factors (``Abelianization'')~\cite{deFlorian:2016gvk}.
Once the two PDFs and the partonic cross sections are evolved from their characteristic scales $m$ and $Q$, respectively, to a common scale $\mu$, the physical cross section in \eq{fact} becomes $\mu$ independent and $\ln^n(m/Q)$ terms are systematically resummed.%
\footnote{The $\mu$ dependence of the physical cross section also cancels at fixed order in $\alpha$, when the corresponding renormalized factorization functions are all evaluated at the same $\mu$ without any RG evolution. In that case the large logarithms are of course only included to the given order in $\alpha$.}
When choosing $\mu\sim Q$ all large logarithms reside in the PDFs.

The DGLAP-type RGEs of the electron (and positron) PDFs have been numerically solved at next-to-leading logarithmic (NLL) order%
\footnote{The RG improved perturbation series of the PDFs is organized as follows: At LL level all terms $\sim \alpha^n L_m^n$ for $n \ge 0$ are resummed, at N$^k$LL all terms $\sim \alpha^{n+r} L_m^n$ with $n \ge 0$, $k\ge r \ge 0$ are included, where  $L_m \equiv \ln(m_e^2/\mu^2)$. The fixed-order expressions at N$^k$LO contain all contributions to $\ord(\alpha^k)$ without any all-order resummation and are part of the N$^k$LL results.}%
, i.e.\ including $\ord(\alpha^2)$ corrections to the corresponding anomalous dimensions (splitting functions) known from QCD, in \rcite{Bertone:2019hks}, which also provides analytical results in the (phenomenologically important) $x \to 1$ limit.
Together with the NLO expressions~\cite{Blumlein:2011mi,Frixione:2019lga}, which serve as the boundary conditions, this RG evolution constitutes the NLL PDFs~\cite{Bertone:2019hks}. For earlier LL accurate results see \rcites{Cacciari:1992pz,Skrzypek:1992vk,Skrzypek:1990qs}.
The resummation of the leading (double) logarithms for the muon (and electron) PDFs in the full SM, i.e.\ taking into account also electroweak and QCD interactions, is carried out and studied in \rcite{Frixione:2023gmf,Garosi:2023bvq,Han:2021kes,Han:2020uid}.
Phenomenological aspects of the NLL electron PDFs as well as their implementation in a Monte Carlo Event generator are discussed in \rcite{Bertone:2022ktl}.
The NNLL DGLAP equations~\cite{Vogt:2004mw,Moch:2004pa} have been solved recursively to $\ord(\alpha^5)$ (and to $\ord(\alpha^6)$ at NLL)
in one-flavor QED in \rcite{Ablinger:2020qvo} (see also \rcite{Arbuzov:2024tac}) using the NNLO PDF results of \rcites{Blumlein:2011mi,Ablinger:2020qvo} as the initial (boundary) condition of the evolution. The numerical effects of these beyond-NLL ISR corrections on a number of precision observables in Drell--Yan-type $e^+e^-$ processes have been analyzed in \rcites{Blumlein:2022mrp,Blumlein:2021jdl,Ablinger:2020qvo}.
NNLL precision of QED ISR beyond the one-flavor approximation requires the NNLO corrections to the electron PDFs due to additional fermion flavors, which involve (logarithms of) the ratios $m_e^2/m_f^2$ and are calculated in the present work.
Such high accuracy (or even higher) is required for
precision measurements at future $e^+e^-$ colliders with large luminosity, e.g.\ at the $Z$ peak, in $ZH$ production, and at the $W^+W^-$ and $t \, \bar{t}$ thresholds.

This paper is organized as follows.
In \sec{def} we give the precise definition of the electron PDFs in terms of operator matrix elements in SCET and trace back two sum rules they obey to charge and four-momentum conservation.
In \sec{factorization} we provide a rough account of the SCET derivation of collinear factorization in QED.
Section~\ref{sec:calculation} outlines our NNLO calculation of the bare PDF matrix elements and \sec{renormalization} describes their renormalization.
In \sec{results} we present our NNLO results for the electron PDFs.
We conclude with a short summary and a brief outlook in \sec{conclusions}.

\section{Operator definition}
\label{sec:def}

To conveniently describe the collinear kinematics of the scattering process we employ light-cone coordinates, where an arbitrary Lorentz vector $p^\mu$ is (Sudakov) decomposed into the components $(p^+,p^-,p_\perp)$ which are defined by
\begin{equation}
  p^\mu = p^- n^\mu/2 + p^+ \bar{n}^\mu/2 + p_\perp^\mu\,,
  \label{eq:SudakovDec}
\end{equation}
with $n^2=\bn^2=0$,  $n\cdot \bn=2$ and $p_\perp\! \cdot n = p_\perp \!\cdot \bn =0$.
In the following the (spatial parts of the) light-like vectors $n^\mu$ and $\bn^\mu$ are assumed to point in opposite directions along the beam axis.

Before presenting the definition of the QED PDFs in terms of SCET operator matrix elements, we emphasize that this definition is in fact equivalent to the traditional ($x$-space) definition in full QED, see e.g.~\rcites{Collins:1981uw,Collins:2011zzd} for the QCD analogs, and leads to the same results.
The reason is that at leading power the $n$-collinear particle modes in the PDF describing an $n$-collinear initial state particle do not interact with modes of other kinematic sectors ($\bn$-collinear, soft, \ldots) as explained below.
Hence, the PDF matrix elements in SCET can simply be regarded as boost-invariant full-QED matrix elements formulated in a boosted frame~\cite{Bauer:2000yr}.
For the same reason we can (and will in practice) use full-QED Feynman rules in the calculations of the PDFs.
Nevertheless we shall discuss here the formal SCET definition, because SCET is a natural and economic framework to derive collinear factorization of high-energy scattering processes as described in \sec{factorization}. See \rcites{iain_notes,Becher:2014oda} for pedagogical introductions to SCET.

In SCET the (``parton-in-electron'') PDF describing a parton $i$ ($=f, \bar{f}, \gamma$), which eventually undergoes a hard scattering process, inside an electron ($e$) is given by
\begin{align} \label{eq:f_def}
f_{i/e}(x) &= \Mae{e^-_n(p^-)}{\oq_i(x p^-)}{e^-_n(p^-)}_\text{av}
\equiv \frac12 \sum_{s=1,2} \Mae{e^-_n(p^-,s)}{\oq_i(x p^-)}{e^-_n(p^-,s)}
\,.\end{align}
This definition applies at the bare and renormalized level depending on whether the composite SCET operator $\oq_i$ is bare or renormalized.
In the latter case the $\oq_i$ and thus the PDF depend on the renormalization scale $\mu$, see \sec{renormalization}.
The external states in \eq{f_def} represent an incoming electron (in direction $n$) with momentum $p^\mu = p^- n^\mu/2 + (m_e^2/p^-) \bn^\mu/2$.
The variable $s$ denotes the electron's spin degree of freedom, which is averaged over for unpolarized beams as indicated by the subscript ``av'' on the matrix element.
For the corresponding parton-in-positron or parton-in-photon PDFs the external states
 have to be replaced accordingly.
The bare fermion ($f$), antifermion ($\bar{f}$), and photon ($\gamma$) PDF operators are defined in terms of collinear SCET fields as~\cite{Bauer:2002nz,Stewart:2010qs}
\begin{align}
\mathcal{Q}^\bare_f(\w)
&= \theta(\w)\, \bar{\chi}^f_n(0) \frac{\bnslash}{2} \bigl[\delta(\w - \bnP_n) \chi^f_n(0)\bigr]
\,, \label{eq:Oferm} \\
\mathcal{Q}^\bare_{\bar{f}}(\w)
&= \theta(\w)\, \tr \Bigl\{\frac{\bnslash}{2} \chi^f_n(0) \bigl[\delta(\w - \bnP_n) \bar\chi^f_n(0)\bigr] \Bigr\}
\,, \label{eq:Oantiferm} \\
\mathcal{Q}^\bare_\gamma(\w)
& = -\w\,\theta(\w)\, \cB_{n\perp\mu}(0) \bigl[\delta(\w - \bnP_n) \cB_{n\perp}^{\mu}(0) \bigr]
\,,
\label{eq:Ophoton}
\end{align}
where the label momentum operator $\bnP_n \equiv \bn \cdot \mathcal{P}_n$~\cite{Bauer:2001ct} acts exclusively on the fields inside the squared brackets in the way described below.
The $n$-collinear PDF operators only involve $n$-collinear SCET field operators. Other momentum modes of particle fields decouple at leading power in SCET.
For the (ultra-)soft modes this can be made manifest by a field redefinition of the collinear fields~\cite{Bauer:2001yt}, which is understood in \eqsm{Oferm}{Ophoton}.


The operators $\chi^f_n$ and $\cB_{n\perp\mu}$ represent gauge-invariant combinations of partonic SCET fields involving a Wilson line $W^{f}_n$ (with $W_n \equiv W^{e}_n$)~\cite{Bauer:2000yr,Bauer:2001ct}:
\begin{align}
  \chi^f_n(y) &= [W_n^f(y)]^\dagger\, \xi^f_n(y)
  \,, \nn\\
  \cB_{n\perp}^\mu(y) &= \frac{1}{e} \bigl[W_n^\dagger(y)\, \ri D_{n\perp}^\mu W_n(y) \bigr]
  = A_{n\perp}^\mu(y) + \frac{1}{e} \bigl[W_n^\dagger(y)\, \cP_{n\perp}^\mu W_n(y) \bigr]
  \,.
  \label{eq:chiBperp}
\end{align}
The operator $\xi_n^f$ represents the $n$-collinear fermion field with flavor $f$ and $\ri D_{n\perp}^\mu = \mathcal{P}_{n\perp}^\mu + e A_{n\perp}^\mu$ is the covariant derivative involving the $n$-collinear gluon field $A_{n}^\mu$,
where $e = \sqrt{4 \pi \alpha}$ is the gauge coupling.
Again, $\mathcal{P}_n$ only acts on the fields inside the squared brackets.
In the following we will often suppress the superscript indicating the flavor on the fermion fields for brevity of presentation.

In the SCET label formalism\footnote{Alternatively, SCET can be formulated entirely in position space, see \rcite{Becher:2014oda} for a comparison.} the collinear fields appearing in \eq{chiBperp} are written as a sum of momentum modes with discrete collinear label momenta ($\tilde{p}$) (excluding the ``zero-bin'' at $\tilde{p} = 0$), while their position ($y$) dependence is associated with their residual ultrasoft momentum components:%
\footnote{In actual momentum space calculations the momentum labels are usually recombined with the ultrasoft momentum components to form continuous momentum variables in practice, see e.g.~\rcites{Stewart:2010qs,iain_notes}.}
\begin{align}
 \xi_n(y) = \sum_{\tilde{p} \neq 0} \xi_{n,\tilde{p}}(y)\,,
\qquad
 A^\mu_n(y) = \sum_{\tilde{p} \neq 0} A^\mu_{n,\tilde{p}}(y) \,.
\end{align}
The fields $\phi_{n,\tilde{p}}$ (with $\phi=\xi,\,A^\mu$) are defined such that for $\bn \cdot \tilde{p} >0$ the operator $\phi_{n,\tilde{p}}$ ($\phi_{n,\tilde{p}}^\dagger$) annihilates (creates) a particle mode, while for $\bn \cdot \tilde{p} < 0$ it creates (annihilates) an antiparticle mode.
Note that for the photon field this implies $[A^\mu_{n,\tilde{p}}(x)]^\dagger = A^\mu_{n,-\tilde{p}}(x)$.
The appropriate label momentum operator $\mathcal{P}_n^\mu$ yields $\tilde{p}^\mu$ when acting on a field $\phi_{n,\tilde{p}}$ and $-\tilde{p}^\mu$ when acting on a field $\phi_{n,\tilde{p}}^\dagger$. When acting on products of fields  $\mathcal{P}_n^\mu$ obeys the Leibniz (product) rule in analogy to ordinary derivatives~\cite{Bauer:2001ct}.
Thus, the delta functions in \eqsm{Oferm}{Ophoton} fix the sum of the label momenta of the parton fields inside $\chi_n$, $\bar{\chi}_n$ and $\cB_{n\perp}^\mu$, respectively.
Of course all interaction vertices in SCET obey label momentum conservation.

In label momentum space the appropriate $n$-collinear QED Wilson line reads%
\footnote{The position space version of $W^f_n(0)$ is $\exp[\ri Q_f e \int_{-\infty}^z \rd s \,\bn\!\cdot\!\mathcal{A}_n(\bn s) ]$, where $z$ is the Fourier conjugate coordinate to the large minus label momentum, and $\mathcal{A}_n(z)$ is the Fourier transform of $A_{n,\tilde{p}}(0)$ w.r.t.\ $\bn\cdot\tilde{p}$~\cite{Bauer:2001yt}.}
\begin{align} \label{eq:Wn}
  W^f_n(y) = \biggl[\, \sum_\mathrm{perms} \exp\biggl(-\frac{Q_f\, e}{\bnP_n}\,\bn \!\cdot\! A_n(y)\biggr)\biggr]\,,
\end{align}
where $Q_{f}$ denotes the electric charge of the fermion $f$ in units of the absolute value of the electron charge $e$ ($Q_e \equiv -1$) and the sum is over all possible attachments of photons to the Wilson line~\cite{Bauer:2000yr,Bauer:2001ct}.
The Wilson line is a unitary operator, so $[W^f_n(y)]^\dagger W^f_n(y) = 1$.
In the SCET power counting both $\bn\!\cdot A_n$ and $\bnP_n$ acting on the $A_n^\mu$ (inside the squared brackets) scale like $Q$. Hence, factors of $W_n$ do not affect the power counting of composite SCET operators.
Note that the last term in \eq{chiBperp} just like the collinear Wilson line involves only unphysical polarizations of the photon field.
In the fermion and antifermion PDF operators $W_n$ represents the leading power remnant of interactions between $n$-collinear photons and $\bn$- and $n_i$-collinear colored particles where $n \!\cdot\! n_i \sim 1$.
From the perspective of the $n$-collinear particles the $n_i$-collinear particles are boosted in $\bn$ direction. Interactions with them can therefore collectively be described at leading power in the EFT by a single eikonal $\bn$-collinear source of the opposite charge of the $n$-collinear fermion that enters the hard scattering process due to charge conservation.
This charge source corresponds to the Wilson line $(W_n^{f})^\dagger$ in \eq{chiBperp}.

The antifermion PDF operator can also be rewritten as
\begin{align}
  \mathcal{Q}^\bare_{\bar{f}}(\w)
  &= \theta(\w)\, \tr \Bigl\{\frac{\bnslash}{2} \bigl[\delta(-\w - \bnP_n) \chi^f_n(0) \bigr] \bar\chi^f_n(0) \Bigr\} \nn\\
  &= - \theta(\w) \bar\chi^f_n(0) \frac{\bnslash}{2} \bigl[\delta(-\w - \bnP_n) \chi^f_n(0) \bigr]\,,
  \label{eq:Oantiferm2}
\end{align}
where the overall minus sign in the second line comes from the anti-commutation of the fermion fields (and an irrelevant infinite constant has been dropped).
For PDFs $\tilde{f}_{i,e}$ defined without the theta functions in \eqsm{Oferm}{Ophoton}, thus permitting negative values of their arguments, \eqs{Oferm}{Oantiferm2} imply $\tilde{f}_{\bar{f}/e}(x) = - \tilde{f}_{f/e}(-x)$, a well-known statement in accordance with the classic QCD literature, see e.g.~\rcite{Collins:1981uw}.
For later convenience we, however, choose to keep the theta functions in \eqs{Oferm}{Oantiferm2} and stick with the PDF definition in \eq{f_def}.

Using \eq{Oantiferm2} we can derive the following sum rule reflecting the conservation of charge (or fermion number) in QED:
\begin{align}
  \int \! \rd x \Big[f_{f/e}(x) - f_{\bar{f}/e}(x)\Big] &=
  \int_{-\infty}^\infty \!\! \rd x \,
  \Mae{e^-_n(p^-)}{\bar\chi^f_n(0) \frac{\bnslash}{2} \bigl[\delta(x p^- - \bnP_n) \chi^f_n(0) \bigr]}{e^-_n(p^-)}_\mathrm{av} \nn\\
  &=\frac{1}{p^-} \,
  \Mae{e^-_n(p^-)}{\bar\chi^f_n(0) \frac{\bnslash}{2} \chi^f_n(0) }{e^-_n(p^-)}_\mathrm{av} \nn\\
  &=\frac{1}{p^-} \,
  \Mae{e^-_n(p^-)}{\bar\xi^f_n(0) \frac{\bnslash}{2} \xi^f_n(0) }{e^-_n(p^-)}_\mathrm{av}  \nn\\
  &=\frac{1}{2p^-} \sum_s \,
  \Mae{e^-_n(p^-,s)}{\bar\psi_f(0) \frac{\bnslash}{2} \psi_f(0) }{e^-_n(p^-,s)}  \nn\\
  &=\frac{\delta_{fe}}{2p^-} \sum_s
  \bar{u}_s(p^-) \frac{\bnslash}{2} u_s(p^-)
  = \frac{\delta_{fe}}{2p^-} \tr \Bigl[ \Bigl(  \frac{\nslash}{2} p^- \!+m \Bigr) \frac{\bnslash}{2}  \Bigr] = \delta_{fe} \,.
  \label{eq:chargesumrule1}
\end{align}
In the fourth line we replaced the SCET fields in terms of full QCD fields as $\xi^f_n(0) =  \frac{\nslash \bnslash}{4} \psi_f(0)$~\cite{Bauer:2000yr,iain_notes}.
The fifth equality implies LSZ wave function renormalization of the electron field. This allows us to use the usual Feynman rule for the external QCD spinors.

Defining the perturbative expansion in the (depending on the context bare or renormalized) coupling of the PDFs as
\begin{align}
  f_{i/j} = \sum_{n=0}^{\infty} \Bigl(\frac{\alpha}{2\pi}\Bigr)^{\!n} f_{i/j}^{(n)} \,,
\end{align}
the charge conservation sum rule of \eq{chargesumrule1}
can be written as\\
\begin{align}
  \int \! \rd x \, \sum_f \Big[ Q_f\, f_{f/e}(x)  + (-Q_f)  f_{\bar{f}/e}(x) \Big] &= Q_e  \label{eq:chargesumrule2} \\
  \Rightarrow \quad   \int \! \rd x \, \Big[ f^{(n)}_{f/e}(x)  -  f^{(n)}_{\bar{f}/e}(x) \Big]  & = 0 \qquad \forall \,n \ge 1\,.
\end{align}

Furthermore, the PDFs are subject to another sum rule originating from energy-momentum conservation:\\
\begin{align}
  \int \! \rd x \, x \, \Big[ \sum_f \big[ f_{f/e}(x)  + f_{\bar{f}/e}(x) \big]
  + f_{\gamma/e}(x) \Big] &= 1 \,.
  \label{eq:momsumrule}
\end{align}
In one-flavor QED, i.e.\ assuming $N_f = 1$ for the number of fermion flavors, this implies
\begin{align}
  \Rightarrow \quad    \int \! \rd x \, x \Big[ f^{(n)}_{e/e}(x)
  + f^{(n)}_{\bar{e}/e}(x)+ f^{(n)}_{\gamma/e}(x) \Big]_{Q_{f\neq e} \to 0}  & = 0 \qquad \forall n \ge 1\,.
 \label{eq:momsumruleNf1}
\end{align}
At the level of operator matrix elements \eq{momsumrule} can be understood as follows.
Rewriting e.g.\ the fermion-in-electron PDF by inserting a complete set of (collinear final) states we have
\begin{align}
  \int_0^\infty \! \rd x \, x\, f_{f/e}(x) &=
  \int_0^\infty \! \rd x \;
  x\, \Mae{e^-_n(p^-)}{
  [\chi^f_n(0)]^\dagger \frac{\nslash \bnslash}{4} \bigl[\delta(x p^- \!- \bnP_n) \chi^f_n(0)\bigr]
  }{e^-_n(p^-)}_\text{av}
  \nn\\
  &=
  \frac{1}{(p^-)^2}\, \Mae{e^-_n(p^-)}{
  [\chi^f_n(0)]^\dagger  \bigl[\, \bnP_n \chi^f_n(0)\bigr]
  }{e^-_n(p^-)}_\text{av}
  \nn\\
  &= \frac{1}{(p^-)^2} \, \sumintX \,
   \Mae{e^-_n(p^-)}{
  [\chi^f_n(0)]^\dagger}{X} \Mae{X}{  \bigl[\, \bnP_n \chi^f_n(0)\bigr]
  }{e^-_n(p^-)} \Big|_\text{av}
\nn\\
  &= \sumintX \; \frac{p^- \!- p_X^-}{p^-} \, \frac{1}{p^-} \Big| \Mae{X}{\chi^f_n(0)
  }{e^-_n(p^-)} \Big|^2_\text{av}
  \,,
  \label{eq:Intxf}
\end{align}
where we employed momentum conservation in order to evaluate the label momentum operator acting on $\chi^f_n$ in the third line.
Note that $(p^- \!-p_X^-)$ is the large longitudinal momentum component carried by the parton $f$ into the hard scattering process in the presence of collinear ISR emission $X$ off the electron. This momentum is weighted in \eq{Intxf} by the squared matrix element for the transition of the electron to a parton $f$ with that momentum accompanied by the real emissions $X$.
Analogous expressions hold for the other PDFs. They are normalized such that $\int \rd x\,x\, f^\zero_{i/j}(x) = \delta_{ij}$.
Accounting for all possibilities of all kinds of partons to carry away fractions of the electron momentum it is clear that the expectation value for the total momentum of all partons equals the electron momentum and we end up with the sum rule in \eq{momsumrule}.%
\footnote{For a detailed account on the corresponding interpretation of the PDFs as parton number densities we refer to \rcite{Collins:2011zzd}.}

The sum rules in \eqs{chargesumrule1}{momsumrule} hold at the bare as well as the renormalized level (after LSZ wave function renormalization), as can be seen from how the PDF operators are renormalized, namely by a Mellin convolution with a renormalization factor, see \sec{renormalization}.
We will use the sum rules for consistency checks of our NNLO calculation in \sec{calculation}.

\section{Collinear factorization}
\label{sec:factorization}

In this section we briefly review the main elements of a proof of collinear factorization in QED within the framework of SCET.
For further details we refer to the quoted literature.
The traditional diagrammatic approach to collinear factorization in full QCD is reviewed e.g.\ in \rcites{Collins:1989gx,Collins:2011zzd}.
We consider the case of cross sections of the type of \eq{fact} where all scales are of $\ord(Q)$ except for the light fermion masses $m_f \sim m \ll Q$.
Due to the conceptual similarity to collinear ISR we also allow for collinear fragmentation of final state partons in the following discussion (albeit in less detail). This will lead to  fragmentation functions besides the PDFs in the factorized cross section.
The SCET power counting parameter is defined by
\begin{equation}
  \lambda = \frac{m}{Q} \ll 1\,.
\end{equation}
The relevant momentum regions of the (loop or phase-space) integrals occurring in the perturbative expansion of the full-theory cross section are identified by a (heuristic) method-of-region~\cite{Beneke:1997zp} analysis, see \rcite{Ma:2025emu} for a recent review.
A priori, for the scattering processes of interest we have to consider the regions with the following momentum scalings%
\footnote{
There may be further regions contributing to individual Feynman diagrams, e.g.\ the ultra-collinear region $(p^+,p^-,p_\perp) \!\sim\! (\lambda^2,1,\lambda)\lambda^2 Q$. In scattering amplitudes, however, their contributions cancel~\cite{terHoeve:2023ehm,Becher:2007cu,Smirnov:1999bza}.
}
\begin{alignat}{2}
  (p^+,p^-,p_\perp) &\sim (1,1,1)Q  \qquad&&\text{(``hard'')}\,, \nn\\
  (p^+,p^-,p_\perp) &\sim (\lambda^2,1,\lambda)Q \qquad &&\text{(``$n$-collinear'')}\,, \nn\\
  (p^+,p^-,p_\perp) &\sim (1,\lambda^2,\lambda)Q \qquad &&\text{(``$\bn$-collinear'')}\,, \nn\\
  (p^+,p^-,p_\perp) &\sim (\lambda,\lambda,\lambda)Q \qquad&&\text{(``soft'')}\,, \nn\\
  (p^+,p^-,p_\perp) &\sim (\lambda^2,\lambda^2,\lambda^2)Q \qquad&&\text{(``ultrasoft'')}\,, \nn\\
  (p^+,p^-,p_\perp) &\sim (\lambda^a,\lambda^b,\lambda)Q\quad
  \qquad&&\text{(``Glauber'')}\,,
  \label{eq:regions}
\end{alignat}
where the exponents for the Glauber scaling are $\{a,b\}=\{2,2\},\{1,2\},\{2,1\}$ and the momentum components $(p^+,p^-,p_\perp)$ are defined according to \eq{SudakovDec}.
In case there are final-state partons that cause logarithmic sensitivity of the cross section to the scale $m$, e.g.~when the measurement fixes their energies ($\sim Q$), one has to allow for corresponding $n_i$-collinear momentum regions (with $n_i \!\cdot\! n \sim n_i \!\cdot\! \bn \sim n_i \!\cdot\! n_j \sim 1$)  in addition to the $n$- and $\bn$-collinear regions associated with the initial-state partons. In the following we will refer to this kind of final-state partons and the initial-state particles commonly as ``resolved''.
There will also be additional Glauber regions with a momentum scaling similar to that in \eq{regions}, but with the perpendicular ($\perp$) component defined relative to a pair of collinear light-cone directions involving at least one $n_i$.

In the following we interpret phase space integrals as cut loop integrals and treat them on the same footing.
Partonic momentum modes that are responsible for loop corrections from the hard region have a typical virtuality of $\ord(Q^2)$ and are integrated out. The corresponding short distance ($\sim 1/Q$) interactions are described by local operators in SCET, where the locality is w.r.t.\ distances $\sim Q/m^2 \gg 1/Q$ conjugate to the residual ultrasoft momenta.
Among the momentum modes with lower virtuality only the resonant (propagating) degrees of freedom, i.e.\ the collinear and (ultra)soft ones, are promoted to fields in SCET and can appear as on-shell particles.

The non-resonant (off-shell) Glauber modes are integrated out. The interactions mediated by Glauber modes are reproduced in SCET by corresponding Glauber operators~\cite{Rothstein:2016bsq} that can couple different ($n$-, $\bn$-, $n_i$-) collinear partons as well as soft and collinear partons.
Hence, Glauber interactions may in general violate the leading-power factorization of the cross section into separate independent factorization functions of hard, collinear, or (ultra)soft origin~\cite{Rothstein:2016bsq,Gaunt:2014ska,Schwartz:2018obd}.
For our problem of describing collinear ISR (and FSR) in QED, however, no (leading-power) factorization violation due to Glauber interactions occurs, because the measurement is by definition assumed to be sufficiently inclusive, i.e.\ operating at scales $\sim Q$ such that the phase space of soft emissions is effectively unconstrained and Glauber exchange is not probed.
In fact, due to the absence of a soft function (and corresponding zero-bin subtractions~\cite{Manohar:2006nz} for the collinear propagators), see below, we can safely neglect Glauber interactions at leading power, cf.~\rcite{Rothstein:2016bsq}.
In any case, even for exclusive processes factorization violation due to Glauber exchange is supposed to be less severe (i.e.~in general appears at higher orders in the coupling) in QED than in QCD, due to the Abelian nature of the interactions.
There is e.g.\ no QED version of the Lipatov vertex, which plays a prominent role in factorization breaking in QCD~\cite{Rothstein:2016bsq,Schwartz:2018obd}.
Such a vertex would couple two Glauber photons to one soft photon, which is not possible in QED, not even at higher loops (according to Furry's theorem).

Furthermore, to describe the effects of collinear ISR and final-state radiation (FSR) in QED with massive fermions for a sufficiently inclusive process at leading power ultrasoft modes are unnecessary.%
\footnote{Note that fermions with mass $\sim m \sim \lambda Q$ are anyway excluded as ultrasoft SCET fields, as they cannot become on-shell.}
The ultrasoft fields not only decouple from the collinear and soft fields in the leading-power SCET Lagrangian upon the BPS field redefinition~\cite{Bauer:2001yt}, corrections from ultrasoft photon loops also vanish as scaleless integrals, because there is no ultrasoft invariant mass scale ($\sim \lambda^2 Q$) in the problem.
The proper EFT framework to derive QED collinear factorization is therefore \SCETII~\cite{Bauer:2002aj,Leibovich:2003jd} involving only soft and collinear fields.

\begin{figure}[t]
  \begin{center}
    \includegraphics[height=0.14 \textheight]{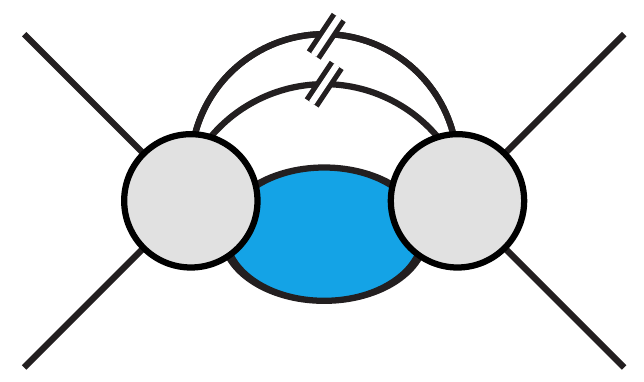}%
    \put(-75,54){$\vdots$}
    \qquad
    \raisebox{38 pt}{$\overset{!}{=}$\;}
    \qquad
    \includegraphics[height=0.15\textheight]{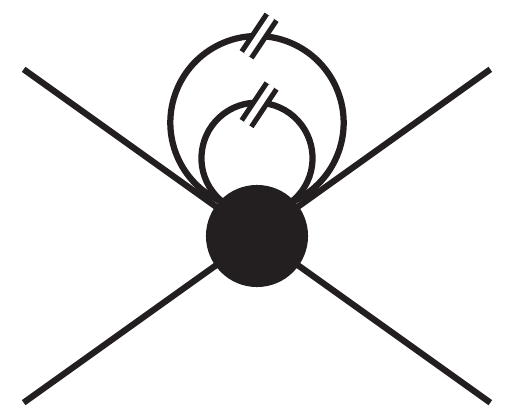}%
    \put(-59,55){$\vdots$}
    \put(-123,78){$\bn$}
    \put(0,78){$\bn$}
    \put(0,0){$n$}
    \put(-123,0){$n$}
    \put(-71,91){$n_1$}
    \put(-71,75){$n_2$}
    \put(-62,17){$C$}
    \quad
    \raisebox{38 pt}{$\;+\;\; \ord(\lambda)$}
  \end{center}
  \caption{Illustration of the hard matching calculation to determine the Wilson coefficient ($C$) of the hard-scattering operator in SCET.
    The LHS represents a squared full-QED matrix element where the (hard) integration over the full phase space of an arbitrary number of unresolved partons (symbolized by the blue ellipse) is carried out.
    The closed lines crossing the final state cut (indicated by the little tilted double lines) correspond to fragmenting partons with fixed momenta ($\sim Q$ in directions $n_i$).
    All external and cut lines are on-shell and do not emit (soft or collinear) radiation.
    LO contributions and virtual (loop) corrections to the scattering amplitude and its complex conjugate are depicted as gray circles.
    The RHS represents the corresponding forward-scattering matrix element with one insertion of the hard-scattering operator (black circle) at leading power in SCET. The labels ($n$, $\bn$, $n_i$) denote the relevant collinear directions.
    For  leading-power matching the LHS may be evaluated with $m_f=0$ (i.e.\ in the hard region) and thus amounts to our definition of the (bare) massless partonic cross section for a given parton channel, i.e.\ a given set of partons involved in the hard scattering process.
    \label{fig:match}
  }
\end{figure}
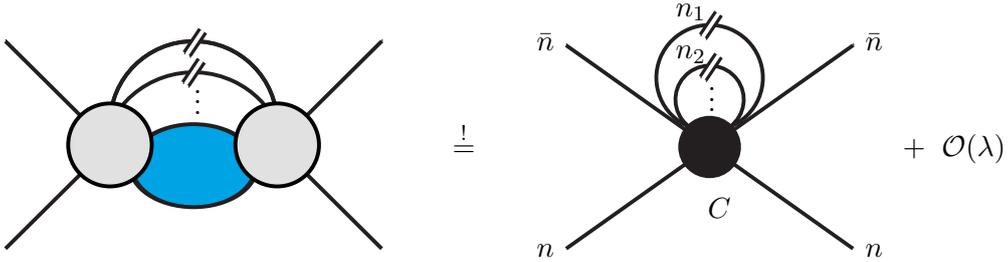

As the first step we determine the Wilson coefficient of the appropriate (squared) hard-scattering operator in SCET by matching to the full-theory process for each partonic channel.
This matching can be conveniently performed at the cross section level, i.e.\ where the hard phase space integrations for all unresolved final-state particles in the full theory have already been carried out~\cite{Bauer:2002nz}.
In this approach the scattered partons that eventually undergo collinear fragmentation into one of the resolved final-state partons can be treated as cut internal lines with fixed on-shell momenta $p_i$ ($p_i\cdot p_j \sim Q \gg p_i^2,p_j^2\sim m$) and are referred to as ``fragmenting'' in the following.%
\footnote{For the matching it makes no difference whether the unconstrained ($\perp$) momentum components ($\sim \lambda Q$) of the cut lines are integrated or not as long as the integrations are preformed on the full-theory as well as on the SCET side.}
On the EFT side the result must be reproduced to leading order in $\lambda$ by  forward-scattering type matrix elements of the hard scattering operator.
The latter is composed of $2(2+N)$ gauge invariant collinear field operators, as defined in \eq{chiBperp}, corresponding to the two ($n$- and $\bn$-collinear) partons initiating the hard scattering and $N$ fragmenting ($n_i$-collinear) partons in the full-theory amplitude and its complex conjugate, respectively.
The sum of large $\ord(Q)$ label momenta ($\omega_j$) along each of the $2+N$ collinear directions is fixed by corresponding delta functions in the operator definition analogous to \eqsm{Oferm}{Ophoton}.
The matching is conveniently performed with exactly $2$ partons in each external state and $N$ cut internal parton lines, see \fig{match}.
Additional leading-power collinear emissions/absorptions from interactions with partons of a different collinear sector and/or hard virtual particles are
fixed by gauge invariance in SCET and incorporated in the Wilson lines of the collinear field operators in \eq{chiBperp}~\cite{Bauer:2000yr,Bauer:2001ct,Bauer:2002nz}.
Collinear interactions within the same sector are described by the collinear SCET Lagrangian~\cite{Bauer:2000yr}.
The outlined matching procedure results in Wilson coefficients of the hard-scattering operators that contain all hard contributions to the process for each partonic channel.%
\footnote{For simplicity of presentation the possibility of different partonic channels, i.e. combinations of initial and fragmenting final-state parton flavors contributing to the partonic cross section and the sum over all those contributions are often implicitly understood in the following.}
Concretely, each coefficient corresponds to the massless ($m_f\to0$) partonic IR-subtracted cross section up to a normalization factor, which amounts to the inverse of the LO SCET matrix element of the respective operator.
In fact, the matching procedure defines the concept of ``partonic cross section'' in the context of processes with QED collinear radiation along the direction of the resolved partons.
In practice, one can determine the Wilson coefficients without evaluating SCET matrix elements. To this end one sets all light fermion masses to zero (which renders the collinear regions scaleless in dimensional regularization) and subtracts the arising collinear IR singularities with the inverse of the (DGLAP-type) renormalization factor responsible for the UV renormalization of the SCET operator, see \sec{renormalization}.

As argued above, the adequate EFT to describe the ``sufficiently inclusive'' massive QED scattering process of interest is \SCETII.
Nonetheless, the matching for the hard-scattering operator proceeds most conveniently via an intermediate matching step to \SCETI{} with only collinear and ultrasoft degrees of freedom and an expansion parameter $\lambda' \equiv \sqrt{\lambda}$~\cite{Bauer:2002aj}.
For the result of the hard Wilson coefficients it does not make any difference whether the hard-scattering operator is part of \SCETI{} or \SCETII.
However, in \SCETI{} one can manifestly decouple ultrasoft and collinear modes in the leading-power Lagrangian via the BPS field redefinition, $\xi_n(x) \to Y_n(x) \xi_n(x)$, $A^\mu_n(x) \to Y_n(x) A^\mu_n(x) Y_n^\dagger(x) = A^\mu_n(x)$~\cite{Bauer:2001yt} (without changing physical predictions), where $Y_n$ is an ultrasoft (position-space) Wilson line in $n^\mu$ direction and $Y_n Y_n^\dagger = 1$.
After the field redefinition ultrasoft and collinear modes and thus also modes from different collinear regimes do not interact with each other anymore.
On the other hand factors of $Y_n^{(\dagger)}$ containing ultrasoft photon fields can now generally appear in the hard-scattering operator.
As an example of such an operator for a process initiated by a fermion and an anti-fermion after the BPS field redefinition consider
\begin{align}
  &\int \!\! \rd \omega_a \rd \omega_a' \rd \omega_b \rd \omega_b' \ldots
  \,C(\omega_a,\omega_a',\omega_b,\omega_b',\ldots)\,
  \overline{\mathrm{T}}\Bigl( \bigl[\overline{\chi}_{n} Y_{n}^\dagger \delta(\omega_a' - \bnP_n^\dagger)\bigr] \overline{\Gamma}^\nu
  \bigl[\delta(\omega_b' - \bnP_{\bn})
  Y_{\bn} \chi_{\bn}\bigr] \ldots \Bigr) \times \nn\\
  &\qquad \mathrm{T} \Bigl(
  \bigl[\overline{\chi}_{\bn}  Y_{\bn}^\dagger \delta(\omega_b - \bnP_{\bn}^\dagger)\bigr] \Gamma^\mu
  \bigl[\delta(\omega_a - \bnP_n) Y_n \chi_n \bigr] \ldots   \Bigr)
,
\label{eq:hardscatop}
\end{align}
where the ellipses may contain further $n_i$-collinear field operators  for fragmenting partons (accompanied by ultrasoft Wilson lines and delta functions fixing their $\ord(Q)$ label momenta from the measurement), $\Gamma^\mu$ denotes a generic Dirac structure ($\overline{\Gamma}^\mu = \gamma_0 \Gamma^{\mu\dagger} \gamma_0 $), and an overall Lorentz tensor contracted with the Lorentz indices $\mu$, $\nu$ is suppressed.
The time- (T) and anti-time-ordered  ($\overline{\mathrm{T}}$) products in round brackets are associated with the full-theory amplitude and its complex conjugate.
The Wilson coefficient $C(\omega_a,\omega_a',\omega_b,\omega_b',\ldots)$ is proportional to the corresponding massless partonic cross section, as noticed above, and a product of label momentum conserving delta functions $\prod_{j=a,b,\ldots} \delta(\omega_j-\omega_j')$, one for each (initial and fragmenting) parton involved in the hard scattering.

The ultrasoft Wilson lines $Y_i^{(\dagger)}$ commute in QED with all fields in \eq{hardscatop} as well as with the label momentum operators (in the delta functions), since they do not carry label momenta.
Since by definition for ``sufficiently inclusive'' processes there are no other (measurement) operators acting on the $Y_i^{(\dagger)}$, it is easy to see that they  cancel pairwise in the hard-scattering operator as  $\overline{\mathrm{T}} [Y_i]\, \mathrm{T}[Y_i^\dagger] = \overline{\mathrm{T}} [Y_i^\dagger]\, \mathrm{T}[Y_i]  = 1$.
After this cancellation the operator is local w.r.t.\ the residual ultrasoft momenta of the collinear SCET fields, i.e.\ they are located at the same point (w.l.o.g.\ $x=0$) in position space, and the (anti-)time ordering can be dropped.
We can now perform the trivial matching of the purely collinear hard-scattering operator in \SCETI{}  (with $\lambda' \equiv \sqrt{\lambda}$) to the respective \SCETII{} operator.
In this way the absence of any soft function contributing to the factorized cross section, cf.~\eq{fact}, is proved.%
\footnote{
Note that matching directly from full QED to \SCETII{} leads to the same result. The derivation of the analog of \eq{hardscatop} with soft Wilson lines is, however, somewhat more cumbersome in that approach~\cite{Bauer:2001yt}.}

The SCET prediction of the cross section is now given by the forward-scattering matrix element with an insertion of the hard-scattering operator (including the sum over all partonic channels),
where the external states contain the two $n$- and $\bn$-collinear particles from the beam.
Since at leading power in SCET there is no cross-talk between the different collinear sectors these external states can be written as a direct product of an $n$- and an $\bn$-collinear one-particle state times $N$ vacuum states of the $n_i$-collinear sectors associated with the resolved final-state partons.
In addition, for each resolved final-state parton $j$ one has to insert
$\sum_{X_{n_j}} |{X_{n_j}} j\rangle \langle {X_{n_j}} j|$
between the two $n_j$-collinear field operators associated with the corresponding fragmenting parton in the hard scattering operator.
The set of states $|{X_{n_j}} j\rangle$ allows for arbitrary unresolved $n_j$-collinear final-state radiation besides the resolved parton $j$.
In a last step one can now (employing Fierz identities) arrange the different collinear components (states, fields, momentum operators) inside the forward-scattering matrix element so as to form separate $n$- and $\bn$-collinear as well as $n_i$-collinear  matrix elements representing PDFs and fragmentation functions, respectively.
For the SCET definition of the latter see \rcite{Procura:2009vm}. The NNLO fermion-initiated fragmentation function in QCD (QED) with one massive flavor was computed in \rcite{Melnikov:2004bm}.
Together with the massless partonic cross sections from the hard Wilson coefficients the PDFs and fragmentation functions constitute the leading-power collinear factorization formula for the cross section, as shown in \eq{fact} for an inclusive final state.
The integrals over the $\omega_j$ in \eq{hardscatop} translate into the convolution integrals over the corresponding longitudinal momentum fractions.

SCET derivations of similar factorization formulas, where many aspects relevant for the derivation sketched here are given in more detail, can be found for instance in \rcites{Fleming:2007qr,Becher:2007ty,Stewart:2009yx}.
We stress that throughout this work SCET is used merely as a tool to establish collinear factorization and to provide a solid theoretical basis for calculating the PDFs. The same results can also be obtained in the traditional full-QED approach.

\section{NNLO calculation of electron PDFs}
\label{sec:calculation}

\begin{figure}[t]
  \begin{center}
    \includegraphics[width=0.23 \textwidth]{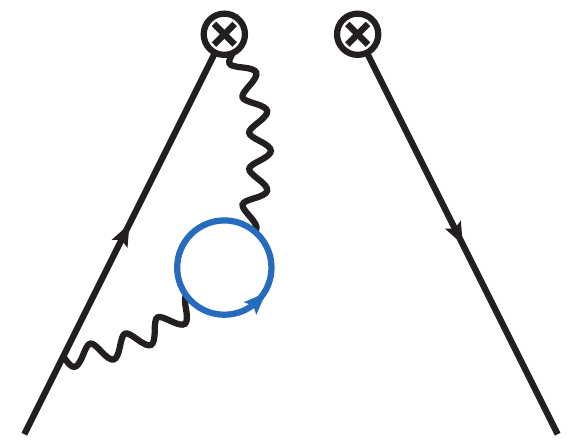}
    \put(-98,65){a)}\;\;
    \includegraphics[width=0.23 \textwidth]{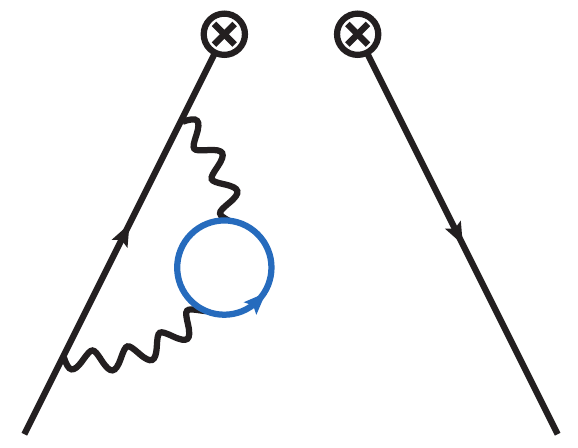}
    \put(-98,65){b)}\;\;
    \includegraphics[width=0.23 \textwidth]{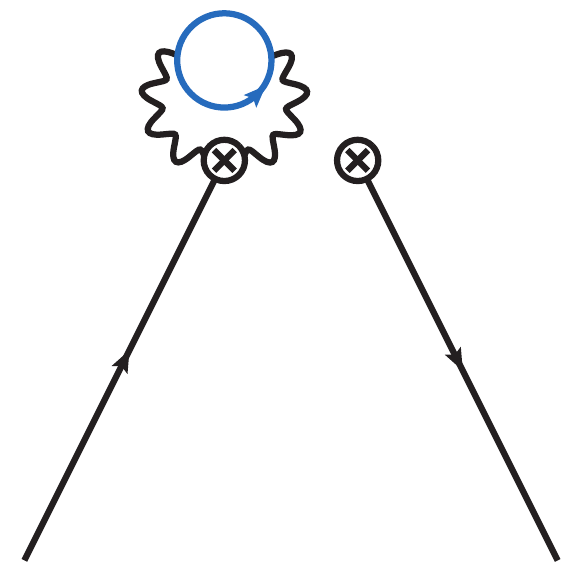}
    \put(-98,65){c)}\;\;
    \includegraphics[width=0.23 \textwidth]{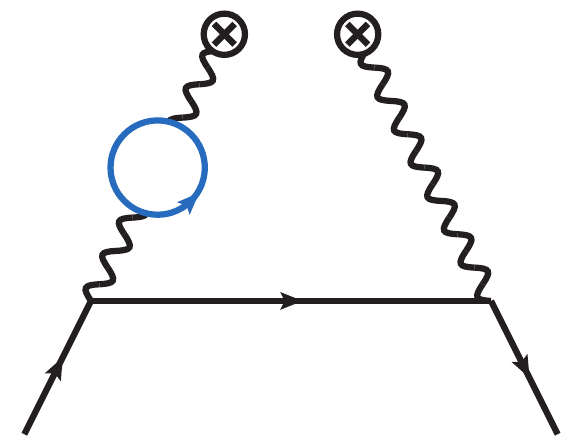}
    \put(-98,65){d)}
    \\[2ex]
    \includegraphics[width=0.23 \textwidth]{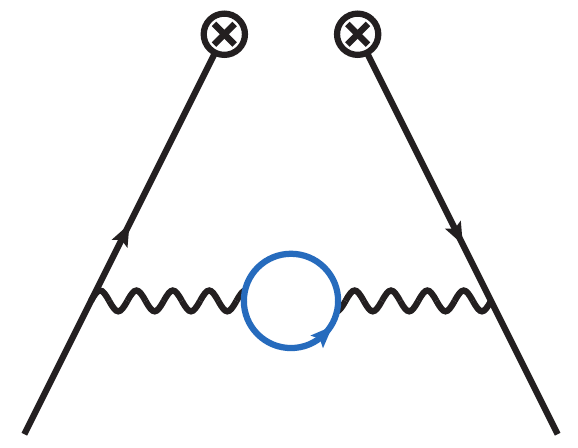}
    \put(-98,65){e)}\;\;
    \includegraphics[width=0.23 \textwidth]{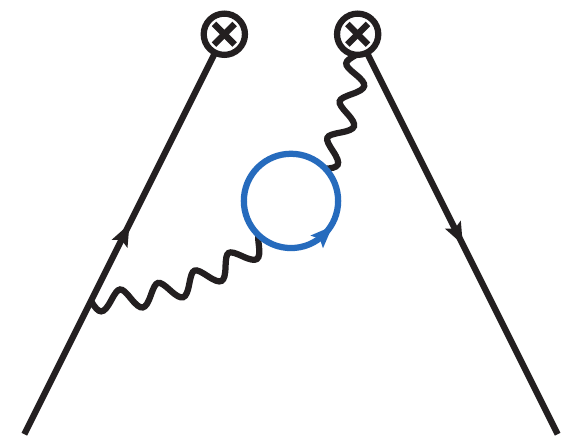}
    \put(-98,65){f)}\;\;
    \includegraphics[width=0.23 \textwidth]{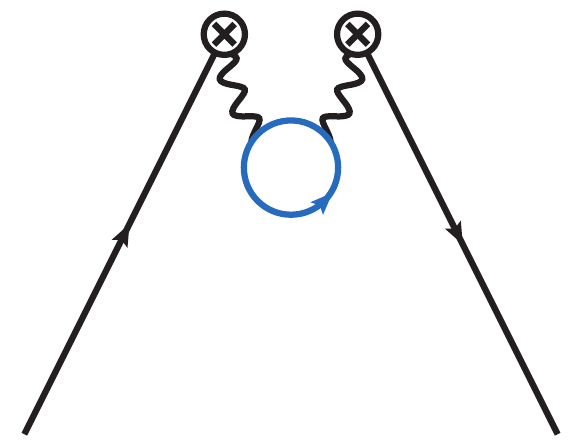}
    \put(-98,65){g)}
  \end{center}
  \caption{Two-loop diagrams with a photon self-energy bubble.
  The black lines represent (w.l.o.g.) massive electrons, the blue lines represent massive fermions of arbitrary flavor.
  The $\otimes$ vertices symbolize the two field operators inside the PDF operator including the collinear Wilson lines according to \eq{chiBperp}, which can directly couple to photons as in diagrams a,c,f,g.
  Note that we draw the $\otimes$ vertices (unlike e.g.\ \rcite{Stewart:2010qs}) with a gap between them, although they represent a local composite operator in SCET.
  This is to emphasize that the PDF can be interpreted in such a way that the two  $\otimes$ vertices lie on the different sides of the final state cut.
  In this sense, diagram~a yields a virtual correction, diagrams~b,c represent wave function renormalization corrections, and diagrams e-g provide real corrections to $f_{e/e}^\two$.
  Diagram d contributes to $f_{\gamma/e}^\two$.
  Left-right mirror graphs (with adapted fermion flow) are not shown, but understood to contribute equally.
  \label{fig:DiagsBub}
  }
\end{figure}

\begin{figure}[t]
  \begin{center}
    \includegraphics[width=0.23 \textwidth]{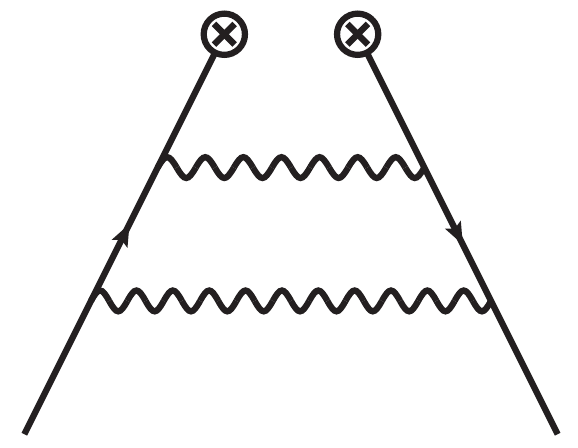}
    \put(-98,65){a)}\;\;
    \includegraphics[width=0.23 \textwidth]{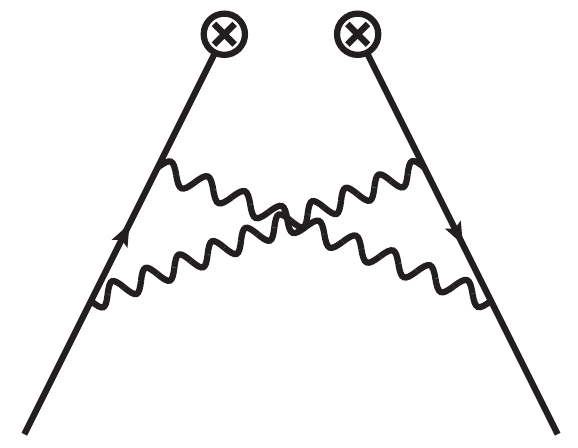}
    \put(-98,65){b)}\;\;
    \includegraphics[width=0.23 \textwidth]{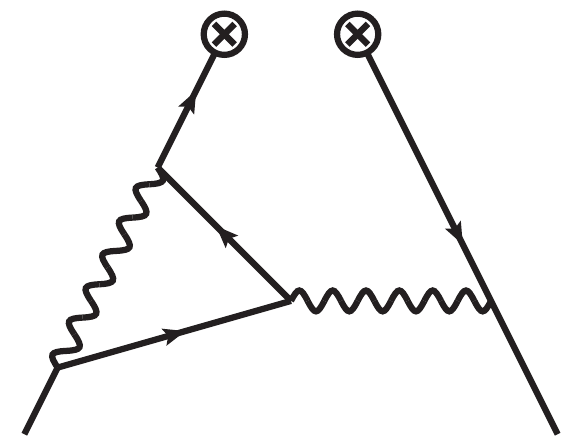}
    \put(-98,65){c)}\;\;
    \includegraphics[width=0.23 \textwidth]{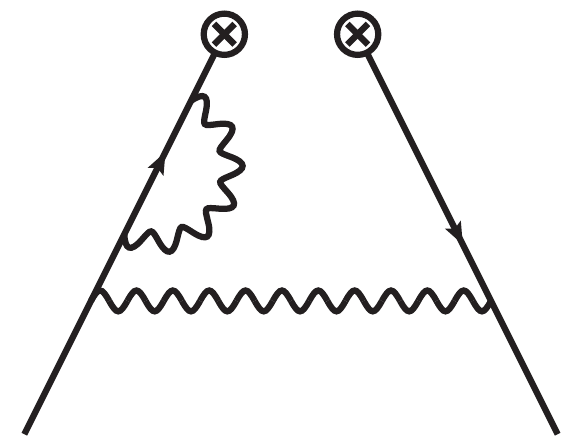}
    \put(-98,65){d)}
    \\[2ex]
    \includegraphics[width=0.23 \textwidth]{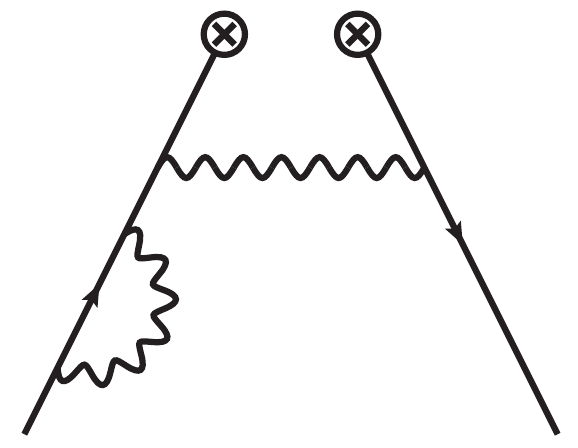}
    \put(-98,65){e)}\;\;
    \includegraphics[width=0.23 \textwidth]{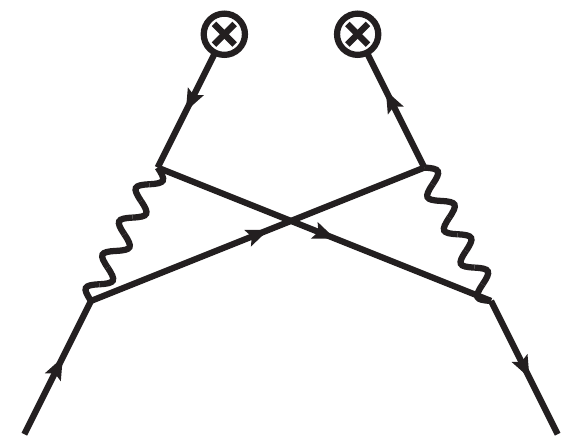}
    \put(-98,65){f)}\;\;
    \includegraphics[width=0.23 \textwidth]{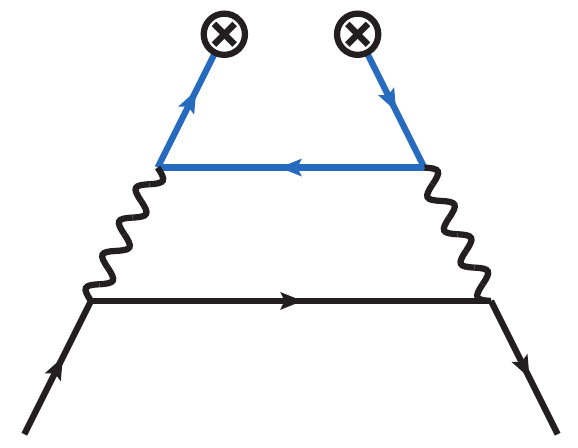}
    \put(-98,65){g)}
  \end{center}
  \caption{Diagrams without photon self energy, but at least one real emission, contributing to $f_{e/e}^\two$ (diagrams\,a-e), $f_{\bar{e}/e}^\two$ (diagram\,f), and $f_{f/e}^\two$ (diagram\,g).
  In contrast to \fig{DiagsBub}, we only show diagrams that are nonzero in $n\cdot A_n=0$ (light-cone) gauge, where the Wilson lines in the PDF operators equal unity.
  In covariant gauges there are many more. Lines and symbols have the same meaning as in \fig{DiagsBub} and we again do not show left-right mirror graphs, which yield equal contributions.
  \label{fig:DiagsReal}
  }
\end{figure}

For the calculation of the bare matrix elements $f_{i/e}$ at NNLO we adopt a similar strategy as employed for the calculation of quark mass corrections to the transverse momentum dependent gluon beam functions in \rcite{Pietrulewicz:2023dxt}, which we outline in the following.
In particular, we exploit the fact that the SCET operators in \eqsm{Oferm}{Ophoton} are
local in time. We can therefore consider the field operators in the PDF matrix elements as time-ordered and compute the higher-order corrections to the PDFs directly in terms of QED loop diagrams involving light-like Wilson lines.%
\footnote{Alternatively one can insert a complete set of (collinear final) states between the two gauge-invariant field operators  ($\chi$, $\overline{\chi}$, $\cB$), interpret the PDF as an integrated squared matrix element, and perform the necessary integrations over the phase space of the emitted collinear partons.}
For details we refer to \rcite{Stewart:2010qs}, where the corresponding one-loop calculation in massless QCD is carried out explicitly.%
\footnote{The calculation in \rcite{Stewart:2010qs} uses SCET Feynman rules, whereas we conveniently use those of full QED, which is equivalent when only $n$-collinear fields are involved~\cite{Bauer:2000yr}.}

At NNLO we have to compute two-loop Feynman diagrams such as the ones in \figs{DiagsBub}{DiagsReal}.
The external electrons are on-shell.
To regulate the UV and IR divergences of individual graphs we employ dimensional regularization with $d=4-2\eps$ spacetime dimensions and the subtraction scale $\tilde\mu \equiv\mu \, e^{\gamma_E/2}(4\pi)^{-1/2}$.
For the purpose of checking the calculation it is useful to distinguish (sub)sets of two-loop diagrams that yield gauge-invariant contributions.
Of course, the sum of all diagrams with a given insertion of one of the PDF operators in \eqsm{Oferm}{Ophoton} is gauge invariant as $f_{i/e}^\two$ is.
Furthermore, the set of purely virtual diagrams as well as the sets of diagrams that admit two-photon, one-photon, and electron-positron final-state cuts, respectively, are separately gauge invariant.
Also, the diagrams with a one-loop gluon self-energy bubble, see \fig{DiagsBub}, form a gauge-invariant subset.
We have performed our calculation in general covariant gauge and confirmed these gauge-invariance statements explicitly by verifying that the respective contributions are independent of the gauge parameter.%
\footnote{In general covariant gauge, the bubble diagrams are individually independent of the gauge parameter, and so are the sets of diagrams with a fermion self energy and the one-loop wave function renormalization correction represented by \fig{DiagsReal}~d and e, respectively}.

To evaluate the diagrams in \fig{DiagsBub} we conveniently express the photon line with the fermion bubble by means of a well-known dispersion relation as a linear combination of a massive photon and a massless photon propagator, see e.g.\ \rcite{Pietrulewicz:2023dxt} for details.
This effectively turns each of the two-loop diagrams in \fig{DiagsBub} into a pair of one-loop diagrams with a massive and a massless photon, respectively.
For the former, in addition to the loop a dispersive integration over the photon mass has to be carried out.
In case of the real-emission diagrams in \fig{DiagsBub}\,e-g these integrations are straightforward in $d$ dimensions even for unequal masses, i.e.\ $m_e \neq m_f$ with fermion $f$ in the loop.
The integrations required for the virtual contribution from diagrams
\ref{fig:DiagsBub}\,a-c, which we conveniently combine at the integrand level%
\footnote{In this way the contributions from the $k^\mu k^\nu$ term of the massive/massless photon propagator ($k^\mu$ being the photon momentum) cancel explicitly in the integrand, cf.\ the related discussion in \rcite{Pietrulewicz:2023dxt}. Note that there are no zero-bin subtractions for the PDF matrix elements, because there is no soft function in the collinear factorization formula.}, is somewhat more involved (for unequal masses), because in contrast to the real-emission diagrams  the minus component of the loop momentum is not fixed by the delta function in \eq{Oferm}.
We still obtain an analytic $\eps$-expanded result by performing one of the loop-momentum component and the photon mass integrations in $d=4$ dimensions after removing non-integrable singularities in the integrand by simple subtraction terms which we integrate in $d$ dimensions.

The diagrams in \fig{DiagsBub}\,a and f are individually rapidity divergent.
To calculate them we employed the rapidity regulator proposed in \rcite{Chiu:2012ir} and isolated the rapidity divergence of diagram \ref{fig:DiagsBub}\,a with a subtraction term.
We find that the rapidity divergences of the two diagrams (and their mirror graphs) cancel exactly leaving a rapidity finite result for the PDF.
This cancellation can be understood as follows.
As full-QCD diagrams are rapidity finite, rapidity divergences of diagrams \ref{fig:DiagsBub}\,a and f are supposed to cancel those of their soft counterparts.
Soft real and virtual corrections, however, cancel exactly (giving rise to a trivial soft function).
Together, this implies the observed cancellation of rapidity divergences between the real diagram \ref{fig:DiagsBub}\,f and the virtual diagram \ref{fig:DiagsBub}\,a.
The diagram in \fig{DiagsBub}\,d contributes to $f_{\gamma/e}^\two$ and represents the only source of an $m_f \neq m_e$ dependence.
In this case we expressed the fermion bubble in terms of an integral over a single Feynman parameter, see e.g.\ \rcite{Pietrulewicz:2023dxt}, which we computed after the loop integrations and the expansion in $\eps$.

We have compared our result for the equal-mass ($m_f = m_e$) contribution to $f_{e/e}^\two$ from the diagrams in \fig{DiagsBub} with the corresponding expression in \rcite{Blumlein:2011mi} and found perfect agreement.
We also exactly reproduced the heavy-flavor loop contribution to the massification factor in \rcite{Wang:2023qbf} for unequal masses ($m_f \neq m_e$).
To this end we had to supplement our calculation of the purely virtual diagrams in \fig{DiagsBub}\,a-c with the two-loop soft virtual contribution, which cancels the rapidity divergence and can be found e.g.\ in \rcite{Pietrulewicz:2017gxc}, see also \rcite{Pietrulewicz:2023dxt}.
The real-emission unequal mass contribution to $f_{e/e}^\two$ is new.

To calculate the remaining contributions to $f_{e/e}^\two$ in general covariant gauge we have to evaluate the diagrams in \fig{DiagsReal}\,a-e as well as all their variants, where the photons are attached to the Wilson lines of the PDF operator in all possible ways. These are the corresponding diagrams, where one or more photon vertices are moved along the electron lines into the  $\otimes$ vertices.
In addition, we need to account for the purely virtual two-loop corrections.
We obtained this piece by multiplying $f_{e/e}^\zero(x) = \delta(1-x)$ with the two-loop massification factor (without fermion loop contributions) given in \rcite{Becher:2007cu}, see also \rcite{Engel:2018fsb}.
There are no purely virtual two-loop contributions to $f_{\bar{e}/e}^\two$ and $f_{\gamma/e}^\two$. These are entirely determined from the diagrams in \fig{DiagsReal}\,f and g, respectively, including their variants with all possible Wilson line attachments.

The modern approach to multi-loop calculations usually includes an integration-by-parts (IBP) reduction of the involved scalar loop integrals to a minimal set of master integrals as a central step.
However, standard (off-the-shelf) IBP reduction of integrals with massive and eikonal (collinear Wilson-line) propagators can be problematic, as it may yield rapidity-divergent master integrals in the reduction of rapidity-finite integrals, see \rcite{Hoang:2019fze} for an explicit example.
Introducing a rapidity regulator even for rapidity-finite integrals leads to well-defined master integrals, but on the other hand increases their number and complexity to an extent that the usefulness of the IBP method is in doubt.
We have therefore adopted a more direct strategy for the calculation of the diagrams in \fig{DiagsReal} and their additional variants in general covariant gauge as outlined in the following.

Using the light-cone decomposition in \eq{SudakovDec} we choose a loop momentum routing such that one loop momentum minus-component is fixed by the delta function in the definition of the PDF operators.
We then carry out the integration over the plus-components of the two loop momenta by residues.
This implies cutting the diagrams in all possible ways and summing the contributions of the different Cutkosky cuts and constrains the momentum fraction $x\le1$ from above. The lower limit $x>0$ is imposed by the Heaviside function in the definition of the PDF operators.
Depending on the cut also the remaining loop momentum minus-component is now constrained to a finite range by kinematics.
Next, we perform the two integrations over the perpendicular loop momenta in $d$ dimensions.
For the more complicated diagrams like the ones in \fig{DiagsReal}\,b, c, and f this requires to introduce a Feynman or Schwinger parameter integral to combine two of the propagator denominators.

We now perform suitable subtractions to render the remaining parameter and loop minus-momentum integrals finite and expand in $\eps$.
The subtraction terms are chosen such that they can be integrated analytically in terms of hypergeometric functions depending on the spacetime dimension $d$. The latter can be expanded in $\eps$ using the \texttt{Mathematica} package \texttt{HypExp}~\cite{Huber:2007dx,Huber:2005yg}.
Finally, we perform the parameter and loop minus-momentum integrations after suitable variable transformations in terms of polylogarithms with the help of the \texttt{Mathematica} package \texttt{PolyLogTools}~\cite{Duhr:2019tlz}, which in turn uses the \texttt{HPL} package~\cite{Maitre:2005uu}.
We regularize $1/(1-x)$ poles in the final expressions by reconstructing a factor $(1-x)^{-1-n\eps}$ with $n=4$ for the diagrams with only a double real cut and $n=2$ for the (sum of) diagrams with a single real cut and re-expanding it in terms of distributions according to \eq{plus_exp}.
Note that this method requires to compute also the $\ord(\eps)$ piece of the unregularized result.

An alternative way to determine the correct distributional structure, at least for the diagrams without eikonal lines, is to promote all singular $\ln^n(1-x)/(1-x)$ terms through $\ord(\eps^0)$ to the corresponding plus distributions $\cL_n(1-x)$, as defined in \eq{plusdef}, and fix the coefficient of $\delta(1-x)$ by a separate calculation of the $x$-integrated diagram.
The latter is conveniently computed by removing any momentum constraints in the PDF operators, i.e.\ the delta and theta functions in \eqsm{Oferm}{Ophoton}, and performing the loop integrations with standard multi-loop technology. Concretely, we used FIRE6~\cite{Smirnov:2019qkx} for the IBP reduction of the two-loop integrals for diagrams \ref{fig:DiagsReal}\,a, b, f, which is safe in the absence of eikonal propagators, and found three master integrals.
The most complicated of these is the equal-mass on-shell sunrise integral, which we took from \rcite{Argeri:2002wz}. The other two master integrals can be easily solved in Feynman parameter representation.
Note that because we removed the theta function of the PDF operator, the fully-integrated diagram \ref{fig:DiagsReal}\,b also contains the result for diagram \ref{fig:DiagsReal}\,f integrated over positive $x$ and vice versa, cf.\ \eq{chargesumrule1}.
The advantage of this method (if applicable) is that we do not need to compute the $\ord(\eps)$ piece of the unregularized result.
We used it to fix the distributional structure of diagram \ref{fig:DiagsReal}\,b analytically (checking it numerically with the other method) and to cross check our expression for diagram \ref{fig:DiagsReal}\,a.

The calculation described in this section yields all bare two-loop ingredients required to determine the PDFs $f_{i/e}$ with $i = e,\bar{e},f,\bar{f}, \gamma$ at NNLO, except for the contribution to $f_{\gamma/e}$ from two-loop diagrams without a closed fermion loop.
That contribution can be obtained from the renormalized NNLO photon-in-electron PDF in one-flavor QED ($N_f=1$), which is computed in \rcite{Ablinger:2020qvo}.
We checked their result for consistency with our renormalized expressions for the other PDFs, which are determined in the next section, by verifying the momentum conservation sum rule in \eq{momsumruleNf1}.

\section{PDF renormalization}
\label{sec:renormalization}

\begin{figure}[t]
  \begin{center}
    \includegraphics[width=0.18 \textwidth]{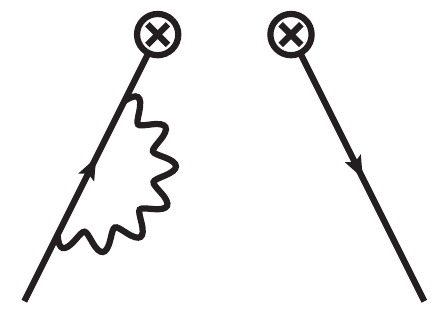}
    \put(-80,45){a)}\qquad\qquad
    \includegraphics[width=0.18 \textwidth]{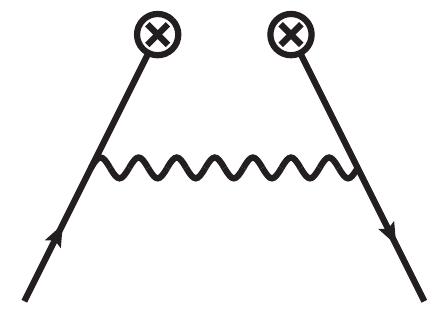}
    \put(-80,45){b)}\qquad\qquad
    \includegraphics[width=0.18 \textwidth]{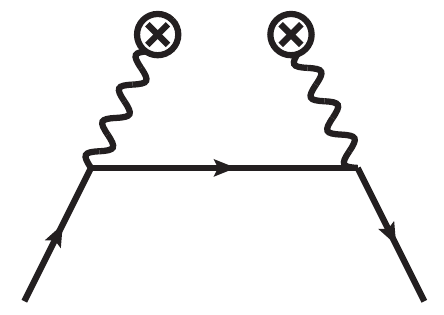}
    \put(-80,45){c)}
  \end{center}
  \caption{One-loop diagrams contributing to $f_{e/e}^\one$ and $f_{\gamma/e}^\one$.
  Graphs with photons attached to Wilson lines are not shown.
  \label{fig:Diags1loop}
  }
\end{figure}

The outcome of the calculation in the previous section are the bare $\ord(\alpha^2)$ PDF matrix elements after wave function renormalization.
Before turning to the NNLO renormalization of the PDF operators we allow for the renormalization of the electron mass $m_e$ and the coupling constant $\alpha$.
To this end we consider the one-loop diagrams contributing to $f_{i/e}$, see \fig{Diags1loop}.
They yield the NLO contributions $f_{e/e}^{\one,\bare}$ and $f_{\gamma/e}^{\one,\bare}$, which we need to $\ord(\eps)$.
The explicit expressions are given in \app{NLOresults}.
The NNLO contribution to the PDFs from $\msb$ coupling renormalization is obtained by replacing the bare $\alpha$ multiplying $f_{e/e}^{\one,\bare}$ and $f_{\gamma/e}^{\one,\bare}$ with the one-loop counterterm%
\footnote{Throughout this paper fermion flavor sums also run over quarks of different colors but the same charge $Q_f$, if present. Each quark flavor therefore comes with a factor $N_c$ for the number of colors.}
\begin{align}
  \delta \alpha \bigl|^\msb =  \frac{\alpha ^2}{3 \pi  \epsilon} \, \sum_f Q_f^2 \,.
  \label{eq:alphaCT}
\end{align}
For the NNLO correction from mass renormalization one has to insert the $\ord(\alpha)$ electron mass counterterm on all internal electron lines of the one-loop diagrams for $f_{e/e}^\one$, so e.g.\ in the diagrams \ref{fig:Diags1loop}\,a, b.%
\footnote{Note that in the massification factor we use to extract the purely virtual corrections to $f_{e/e}^\two$ without fermion loop the mass renormalization contribution is already included.}
In the on-shell scheme this counterterm reads
\begin{equation}
  \delta m_e \big|^{\text{on-shell}} = Q_e^2\, \frac{\alpha}{4 \pi} \,  m_e
  \biggl[
  -\frac{3}{\epsilon }
  +3 L_m -4
  -\biggl(\frac{3}{2} L_m^2
  -4 L_m  +\frac{\pi ^2}{4}+8\biggr) \eps
  +\ord(\eps) \biggr]\,,
  \label{eq:mCT}
\end{equation}
where  $L_m \equiv \ln(m_e^2/\mu^2)$.
On the RHS of \eqs{alphaCT}{mCT} as well as in all other explicit results in the following $\alpha \equiv \alpha(\mu)$ is understood to denote the $\msb$ renormalized QED coupling.
Adding the NNLO coupling and mass renormalization corrections to the results of \sec{calculation} we arrive at the bare PDF expressions $f_{i/e}^{\two,\bare}$, where ``bare'' now only refers to the PDF operators in the matrix elements.
The relation between bare and renormalized PDF operators is~\cite{Stewart:2010qs}
\begin{align} \label{eq:Oop_ren}
  \mathcal{Q}_i^\bare(\w)
  &= \sum_j \int\! \frac{\rd \w'}{\w'}\, Z^f_{ij}\Big(\frac{\w}{\w'},\mu\Big) \mathcal{Q}_j(\w',\mu)\,.
\end{align}
At the level of PDF matrix elements this translates to
\begin{align}
  f^\bare_{i/j}(x) = \sum_k  Z_{ik}(x, \mu) \convx f_{k/j}(x, \mu) \,,
\end{align}
 where we introduced the symbol $\otimes_x$ for the (Mellin-)convolution
\begin{align}
  f(x)\convx g(x)\equiv
  \int \!\!\rd y \int \!\!\rd z\; \delta(x - yz) \,f(y) \, g(z)
  =
  \int \frac{\rd y}{y}f(y) \, g\Big(\frac{x}{y}\Big)
  = g(x)\convx f(x)\,.
\end{align}
The inverse of the PDF renormalization factor $Z_{ij}$ is defined through
\begin{align}
  \sum_k  Z^{-1}_{ik}(x,\mu) \convx Z_{kj}(x,\mu)  \equiv \delta_{ij} \, \delta(1-x)\,,
\end{align}
and for its perturbative expansion we adopt the same convention as for the PDFs:
\begin{align}
  Z_{ij} = \sum_{n=0}^{\infty} \Bigl(\frac{\alpha}{2\pi}\Bigr)^{\!n} Z_{ij}^{(n)} \,.
\end{align}
The renormalized PDFs are thus given by
\begin{align}
f_{i/j}(x,\mu) ={}& \sum_k  Z^{-1}_{ik}(x, \mu) \convx f^\bare_{k/j}(x)
\nn\\
={}& \delta_{ij} \, \delta(1-x) + \frac{\alpha}{2\pi}
\biggl\{ \bigl[Z^{-1}\bigr]^\one_{ij} + f^{\bare,\one}_{i/j} \biggr\}
\nn\\
&{}+ \Bigl(\frac{\alpha}{2\pi}\Bigr)^{\!2}
\biggl\{ \bigl[Z^{-1}\bigr]^\two_{ij} +\sum_k  \bigl[Z^{-1}\bigr]^\one_{ik} \convx f^{\bare,\one}_{k/j} + f^{\bare,\two}_{i/j} \biggr\}
\,+\, \ord(\alpha^3) \,,
\label{eq:NNLOren}
\end{align}
where we suppressed the arguments of the $Z$ factor and bare PDF terms after the second equal sign for brevity.
The renormalization factor $Z_{ij}$ can be expressed in terms of the QED version of the well-known (DGLAP-type) collinear splitting functions $P_{ij}$, which in turn determine the RG evolution of the PDFs:
\begin{align} \label{eq:f_RGE}
\mu\frac{\rd}{\rd\mu} f_{i/j}(x,\mu)
= \sum_k 2 P_{ik} [x,\alpha(\mu)] \convx f_{k/j}(x,\mu) \,.
\end{align}
Explicitly, the perturbative coefficients of the NNLO $Z$ factor in the $\msb$ scheme are
\begin{align}
Z^{\zero}_{ij}(x)
&= \delta_{ij} \, \delta(1-x)
\,, \\
Z^{\one}_{ij}(x)
&=  \frac{1}{\epsilon}P^\zero_{ij}(x)
\,, \\
Z^{\two}_{ij}(x)
&=  \frac{1}{2\epsilon^2} \sum_k P^\zero_{ik}(x) \convx P^\zero_{kj}(x) + \frac{1}{3\epsilon^2}P^\zero_{ij}(x) \sum_f Q_f
+ \frac{1}{2\epsilon}P^\one_{ij}(x)
\,,\end{align}
and we therefore have
\begin{align}
\bigl[Z^{-1}\bigr]^\zero_{ij}(x)
&= \delta_{ij} \, \delta(1-x)
\,, \\
\bigl[Z^{-1}\bigr]^\one_{ij}(x)
&= - \frac{1}{\epsilon}P^\zero_{ij}(x)
\,, \\
\bigl[Z^{-1}\bigr]^\two_{ij}(x)
&= \frac{1}{2\epsilon^2} \sum_k P^\zero_{ik}(x) \convx P^\zero_{kj}(x) - \frac{1}{3\epsilon^2}P^\zero_{ij}(x) \sum_f Q_f
- \frac{1}{2\epsilon}P^\one_{ij}(x)\,.
\label{eq:invZ2}
\end{align}
The QED expressions of the one- and two-loop splitting functions as well as their convolutions required for the NNLO renormalization of $f_{i/e}$ are given in \app{splitting}.

\section{Results}
\label{sec:results}

Using \eq{NNLOren} we obtain the following
NNLO contributions to the renormalized electron structure functions in QED with $N_f$ (massive) flavors:
\begin{align}
  \label{eq:f2ee}
  f^\two_{e/e}(x,\mu) &= Q_e^2 \, \theta(x) \biggl[ Q_e^2\, \varphi^{\two,N_f=1}_{e/e}(x,\mu)
  \;+\; \sum_{f \neq e} Q_f^2 \,  \varphi^{\two,f \neq e}_{e/e}(x,\mu) \biggr] \,, \\
  f^\two_{\bar{e}/e}(x,\mu) &=  Q_e^4 \,  \theta(x) \, \varphi^{\two,N_f=1}_{\bar{e}/e}(x,\mu) \,, \\
  f^\two_{f/e}(x,\mu) &=  f^\two_{\bar{f}/e}(x,\mu)  = Q_e^2  Q_f^2 \, \theta(x)\, \varphi^{\two,f \neq e}_{f/e}(x,\mu) \,, \\
  f^\two_{\gamma/e}(x,\mu) &= Q_e^2 \, \theta(x) \biggl[
  Q_e^2\, \varphi^{\two,N_f=1}_{\gamma/e}(x,\mu)
  \;+\; \sum_{f \neq e}  Q_f^2 \, \varphi^{\two,f \neq e}_{\gamma/e}(x,\mu) \biggr]\,.
  \label{eq:f2game}
\end{align}
We have split each of the $f^\two_{i/e}$ expressions into a contribution
that equals the result in one-flavor QED (with only photons and electrons) and the contribution that arises due to the presence of further flavors $f\neq e$. They are indicated by the superscripts $N_f=1$ and $f\neq e$, respectively.
Our results for the functions $\varphi^{\two,N_f=1}_{e/e}$ and $\varphi^{\two,N_f=1}_{\bar{e}/e}$ agree with those of \rcite{Blumlein:2011mi} and read
\begin{align}
\varphi^{\two,N_f=1}_{e/e} &=
-4 (x+1) \ln ^3(1-x)+\left(-6 L_m (x+1)-\frac{9 x^2+1}{2 (x-1)} \ln (x)-10\right) \ln ^2(1-x)
\nn\\
&+ \ln (1-x) \Biggl[-2 (x+1) L_m^2+\frac{1}{3} (-5 x-29)
L_m+\frac{5 \pi ^2 x^2+13 x^2-21 x-3 \pi ^2+8}{2 (x-1)}
\nn\\
&\quad +\ln (x) \left(\frac{2 \left(x^2+1\right)}{x-1} L_m-\frac{4 x^4+7 x^3-38 x^2+29 x+4}{3 (x-1)
  x}\right)+\frac{7 \left(x^2+1\right)}{2 (x-1)} \ln ^2(x)\Biggr]
\nn\\
&  -\frac{x+1}{4}  \ln ^3(x)+\biggl(\frac{7}{2} L_m (x+1)+\frac{8 x^5+87 x^4-208
  x^3+179 x^2-58 x+8}{12 (x-1)^3}\biggr) \!\ln ^2(x)
\nn\\
&  +\biggl[4 L_m^2+\frac{34 L_m}{3}-\frac{4}{3} \left(3+2 \pi ^2\right)\biggr] \cL_1(1-x)+(12
L_m+12) \cL_2(1-x)+8 \cL_3(1-x)
\nn\\
& +\frac{7-13 x^2}{x-1} \Li_3(x)
 +\frac{1-7 x^2}{x-1} \Li_3(1-x)
 +\Li_2\left(x^2\right) \Biggl[-\frac{2
  \left(x^2-3\right)}{x-1} \ln (x)
\nn\\
&\quad  -\frac{4 \left(x^4+2 x^3-3 x^2+5 x+1\right)}{3 (x-1) x}\Biggr]
 +\frac{2 \left(x^2-3\right)}{x-1}
\Li_3\left(x^2\right)
+\Li_2(x) \Biggl[-2 L_m (x+1)
\nn\\
&\quad +\frac{4 x^4+18 x^3+2 x^2-x+4}{3 (x-1) x}+\frac{1-7 x^2}{x-1} \ln (1-x)+\frac{7 x^2-1}{x-1} \ln
(x)\Biggr]
\nn\\
&+\frac{-2 x^3-2 x^2-5 x+2}{3 x} L_m^2+\frac{36 x^3+18 \pi ^2 x^2-239 x^2+18 \pi ^2 x+169 x+36}{18 x} L_m
\nn\\
&+\ln (x) \Biggl[-\frac{1}{18 (x-1)^4 (x+1)^3}
\Bigl(40 x^9+440
  x^8-1059 x^7-747 x^6+2329 x^5-351 x^4
\nn\\
&\qquad
  -297 x^3-97 x^2-117 x+115 \Bigr)
 -\frac{8 \left(x^4+2 x^3-3 x^2+5 x+1\right)}{3 (x-1) x} \ln (x+1)
\nn\\
& \quad   +\frac{5 x^2-1}{2 (x-1)} L_m^2  -\frac{8 x^4-12 x^3-18 x^2+15 x+8}{3 (x-1) x} L_m\Biggr]
  +\cL_0(1-x) \Bigg[\frac{7 L_m^2}{3}
\nn\\
&\quad -\frac{1+12 \pi ^2}{9} L_m -\frac{4(13+9 \pi ^2-108 \zeta_3)}{27} \Bigg]
+\frac{1}{54 (x-1)^3 x  (x+1)^2} \Bigl[400 x^8
\nn\\
&\quad +162 \zeta_3 x^7 +30 \pi ^2 x^7-830 x^7 -30 \pi ^2 x^6-476 x^6+54 \zeta_3 x^5
 -69  \pi ^2 x^5+822 x^5
 \nn\\
&\quad +60 \pi ^2 x^4+1208 x^4  -594 \zeta_3 x^3+48 \pi ^2 x^3-594 x^3-30 \pi ^2 x^2+100 x^2 +378 \zeta_3 x
\nn\\
&\quad -9 \pi ^2 x-358 x+112 \Bigr]
+\delta(1-x)\Biggl[\frac{15-8 \pi ^2}{24} L_m^2+\frac{1}{72} \left(-135+4 \pi ^2+144 \zeta_3\right) L_m
\nn\\
&\quad +\frac{32077+648 \pi
  ^2-360 \pi ^4+18576 \zeta_3}{2592}-\pi ^2 \ln (4)\Biggr]\,,
\label{eq:phieeNf1}
\end{align}

\begin{align}
  \varphi^{\two,N_f=1}_{\bar{e}/e} &=
-\frac{(x-1) \left(4 x^2+7 x+4\right)}{6 x} L_m^2
+\ln (x) \Biggl[-\frac{(x-1) \left(8 x^2+17 x+8\right)}{3 x} L_m
\nn\\
&\quad +\ln (x+1) \Biggl(\frac{4  \left(x^2+1\right)}{x+1} L_m-\frac{2 \left(4 x^4+17 x^3+10 x^2-19 x-4\right)}{3 x (x+1)}\Biggr)
\nn\\
&\quad -\frac{4 \left(x^6+7 x^5+21 x^4+12 x^3-x^2-3 x-1\right)}{3 x
  (x+1)^3} \ln (1-x)+L_m^2 (x+1)
\nn\\
&\quad  -\frac{40 x^4+335 x^3+427 x^2+153 x+69}{18 (x+1)^2}\Biggl]
+\Li_2\bigl(x^2\bigr) \Biggl[\frac{2 \left(x^2+1\right)}{x+1} L_m
\nn\\
&\quad
-\frac{x^2+8 x+1}{x+1} \ln (x)
-\frac{4 x^4+17 x^3+10 x^2-19 x-4}{3 x (x+1)}\Biggr]
\!+\Li_2(x) \Biggl[-\frac{4 \left(x^2+1\right)}{x+1} L_m
\nn\\
&\quad +\frac{8
  x}{x+1} \ln (x)+\frac{2 \left(2 x^6+11 x^5+6 x^4-6 x^3-30 x^2-21 x-2\right)}{3 x (x+1)^3}\Biggr]
\nn\\
&  +\frac{6 x^4+\pi ^2 x^3-3 x^3-12 x^2+\pi ^2 x+3 x+6}{3 x (x+1)}L_m
+\ln ^2(x) \Biggl[ \frac{2 \left(x^2+3 x+1\right)}{x+1} L_m
\nn\\
&\quad +\frac{x^2+1}{x+1} \ln (x+1)+\frac{8 x^5+99 x^4+348 x^3+290 x^2+84 x+3}{12 (x+1)^3}\Biggr]
\nn\\
&  -\frac{12 \left(x^2+1\right)}{x+1}\, \Li_3\biggl(\frac{1}{x+1}\biggr)
-\frac{2 \left(x^2+8 x+1\right)}{x+1}\, \Li_3(x)
+\frac{3 x^2+16 x+3}{2  (x+1)}\, \Li_3\bigl(x^2\bigr)
\nn\\
&  +\frac{2 \left(x^2+1\right)}{x+1} \ln ^3(x+1)-\frac{\pi ^2 \left(x^2+1\right)}{x+1} \ln (x+1)-\frac{x}{3 (x+1)} \ln^3(x)
\nn\\
&+\frac{1}{54 x (x+1)^3} \Bigl[400 x^6+594 x^5\zeta_3+9 \pi ^2 x^5+1326 x^5+1620 x^4\zeta_3+108 \pi ^2 x^4
\nn\\
&\quad+1164 x^4 +2052 x^3\zeta_3+90 \pi ^2 x^3-576 x^3+1620 x^2\zeta_3
  +84  \pi ^2 x^2-1452 x^2
\nn\\
&\quad  +594 x\zeta_3+45 \pi ^2 x-750 x-112 \Bigr] \,.
\end{align}
The plus distributions $\cL_n$ in \eq{phieeNf1} are defined in \eq{plusdef}.
The function $\varphi^{\two,N_f=1}_{\gamma/e}$ was computed in \rcite{Ablinger:2020qvo}.
In our notation the explicit expression is
\begin{align}
  \varphi^{\two,N_f=1}_{\gamma/e} &=
\frac{-11 x^2+28 x-16}{12 x} L_m^2
+\ln (x) \Biggl[\frac{-31 x^2+32 x-32}{6 x} L_m+\frac{1}{2} L_m^2 (2-x)
\nn\\
&\quad -\frac{20 \pi ^2 x^2+231 x^2-40 \pi ^2 x-335
  x+320}{60 x}\Biggr]
  +\ln (1-x) \Biggl[\frac{x^2-2 x+2}{x} L_m^2
\nn\\
&\quad  +\frac{13 x^2-8 x+20}{3 x} L_m-\frac{8 (x-2)^3 }{3 x^2}\ln (2-x)
  -\frac{7 x^3+12  x^2-48 x+32 }{6 x^2}\ln (x)
\nn\\
&\quad  + \frac{1}{90 x^4}  \Bigl(60 \pi ^2 x^5+819 x^5-120 \pi ^2 x^4-1440 x^4+120 \pi ^2 x^3  +1320 x^3  -400 x^2
\nn\\
&\quad -640 x+256 \Bigr)   \Biggr]
  +\Li_2(x) \Biggl[-\frac{4 \left(x^2-2 x+2\right)}{x} \ln (1-x)+2 L_m (x-2)
\nn\\
&\quad  +\frac{13 x^3-36 x^2+48 x-32}{6 x^2}+(x-2) \ln (x)\Biggr]
  -\frac{8 \left(x^2-2 x+2\right)}{x} \Li_3(1-x)
\nn\\
&  +\frac{-6 \pi ^2 x^2+85 x^2+12 \pi ^2 x-5 x+32}{18 x} L_m
  +\ln ^2(1-x) \Biggl[\frac{3 \left(x^2-2 x+2\right)}{x} L_m
\nn\\
&\quad  -\frac{4 \left(x^2-2 x+2\right)}{x} \ln (x)+\frac{41 x^3-118 x^2
  +190  x-96}{12 x^2}\Biggr]
  +\frac{8 (x-2)^3 }{3 x^2}\Li_2(1-x)
\nn\\
&  +\frac{4 \left(x^2-2 x+2\right)}{x}
\Li_3\bigl[(1-x)^2\bigr]+\Li_2\bigl[(1-x)^2\bigr] \Biggl[-\frac{4 \left(x^2-2 x+2\right)}{x} \ln (1-x)
\nn\\
&\quad -\frac{4 (x-2)^3}{3 x^2}\Biggr]
-\frac{x^2-2  x+2}{6 x} \ln ^3(1-x)
+\biggl[\frac{1}{2} L_m (2-x)+\frac{1}{24} (23 x+12)\biggr] \ln ^2(x)
\nn\\
&
+2 (x-2)\Li_3(x) +\frac{1}{12} (2-x) \ln ^3(x)
  -\frac{1}{540 x^3} \Bigl[-1080 x^4 \zeta_3+90 \pi ^2 x^4+2005 x^4
\nn\\
&\quad  +2160 x^3 \zeta_3-720 \pi ^2 x^3+316 x^3-4320 x^2 \zeta_3
  +1440 \pi ^2 x^2-1312 x^2-960 \pi ^2 x
\nn\\
&\quad  +3072 x-1536 \Bigr] \,.
\end{align}
The functions $\varphi^{\two,f\neq e}_{i/e}$ represent new results and are given by
\begin{align}
  \varphi^{\two,f \neq e}_{e/e} &=
  \ln(x) \Biggl[-\frac{2 (x^2+1)}{3 (x-1)} L_m
  -\frac{11 x^2-12 x+11}{9 (x-1)}\Biggr]
 +\ln(r) \Biggl[\frac{2 (x^2+1)}{3 (x-1)} \ln  (x)
 +\frac{8(2 x-1)}{9}
\nn\\
&\quad  -\frac{4}{3} (x+1) \ln(1-x)\Biggr]
  +\frac{4 \left(x^2+1\right)}{3 (x-1)} \ln^2\Bigl(u+\sqrt{u^2+1}\Bigr)
  -\frac{x^2+1}{6 (x-1)} \ln ^2(x)
\nn\\
&  +\delta(1-x) \Biggl[-\frac{L_m^2}{2}+\frac{1}{18} \left(27+4 \pi ^2\right) L_m
   +\frac{2 (27 r-55) }{9r^{3/2}}  \Li_2\bigl(\sqrt{r}\bigr)
\nn\\
&\quad   -\Li_2(r) \Biggl( \frac{\left(\sqrt{r}+1\right) \left(26 r^{3/2}+r-\sqrt{r}-54\right)}{18 r^2} + \frac{2 \ln(r)}{3} \Biggr)
+\frac{4\Li_3(r)}{3}
\nn\\
&\quad   -\ln(r) \Biggl(\frac{(27 r-55) \ln\left(\sqrt{r}+1\right)}{9 r^{3/2}}
  +\frac{\left(\sqrt{r}-1\right) \left(26 r^{3/2}-r-\sqrt{r}+54\right) \ln(1-r)}{18r^2}
\nn\\
&\qquad   +\frac{2 \left(3 \pi ^2 r-13 r+42\right)}{27 r}\Biggr)
   +\frac{-432 r \zeta_3+120 \pi ^2 r-2837 r+5976}{648 r}+\frac{\ln^2(r)}{2}\Biggr]
\nn\\
&   +\cL_0(1-x)
  \Biggl[-\frac{2 L_m^2}{3}+\frac{8 L_m}{9}+\frac{2 \ln^2(r)}{3}-\frac{8 \ln(r)}{9}+\frac{56}{27}\Biggr] +\frac{1}{3} L_m^2 (x+1)
\nn\\
&  +\cL_1(1-x)
  \Biggl[\frac{8 \ln(r)}{3}-\frac{8 L_m}{3}\Biggr] -\frac{8}{9} L_m (2 x-1)
  +\frac{4}{3} L_m (x+1) \ln(1-x)
\nn\\
&  + \ln\Bigl(\sqrt{u^2+1}+u\Bigr)
   \Biggl[\frac{2 \sqrt{u^2+1} \left(5 x^2-12 x+5\right)}{9 u^3 (x-1)}
  -\frac{16 \sqrt{u^2+1} \left(2 x^2-3 x+2\right)}{9 u  (x-1)}\Biggr]
\nn\\
& -\frac{1}{3} (x+1) \ln^2(r)
-\frac{2  \left(5 x^2-12 x+5\right)}{9 u^2 (x-1)}
+\frac{19 x^2-42 x+75}{27 (x-1)}
\,,
  \label{eq:phieeQf}
\end{align}

\begin{align}
  \varphi^{\two,f \neq e}_{f/e} &=
  -\frac{1}{6} (x+1) \ln ^3(x)+\Bigl[ 3 L_m (x+1)+\frac{1}{12} \left(8 x^2+72 r x+15 x+3\right)\Bigr] \ln ^2(x)
\nn\\
&  +\ln (x)\Biggl[ (x+1) L_m^2
     -\frac{8 x^3+3 x^2- 15  x - 8}{3 x} L_m
     +\frac{(x-1) \left(4 x^2+7 x+4\right)}{3 x} \ln (1-x)
\nn\\
&\quad    +(4 r x-x-3) \ln (1-r x)
     -\frac{8 \bigl(x^3+3 r x^2-3 x-1\bigr)}{3 x} \ln \bigl(1-r x^2\bigr)
\nn\\
&\quad -\frac{1}{9 (r x-1) \bigl(r x^2-1\bigr)^3}\Bigl(20 r^4 x^9+150 r^4
  x^8-20 r^3 x^8-96 r^4 x^7-213 r^3 x^7
\nn\\
&\qquad  +36 r^4 x^6-327 r^3 x^6+72 r^2 x^6+216 r^3 x^5+543 r^2 x^5-96 r^3 x^4-63 r^2 x^4
\nn\\
&\qquad -156 r x^4+96 r^2 x^3-191 r x^3+12 r^2 x^2-69 r   x^2+56 x^2-24 r x+33 x+21\Bigr)
\Biggr]
\nn\\
& +\ln ^2(r) \Biggl[\frac{(x-1) \left(4 x^2+7 x+4\right)}{6
    x}-(x+1) \ln (x)\Biggr]
    + \Li_2(x) \Biggl[\frac{(x-1) \left(4 x^2+7 x+4\right)}{3 x}
\nn\\
&\quad    +2 (x+1) \ln (x)\Biggr]
    +\Li_2(r x) \Bigl[4 r x-x+2 (x+1) \ln (r)   +2 (x+1) \ln (x)-3 \Bigr]
\nn\\
&  - \Li_2\bigl(r x^2\bigr) \Biggl[\frac{4 \left(x^3+3 r x^2-3 x-1\right)}{3 x}
  + 2 (x+1) \ln (r) + 4 (x+1) \ln (x)\Biggr]
\nn\\
&  +\frac{(x-1) \left(2 x^2-5 x-2\right)}{x} L_m
  -\frac{(x-1) \left(4 x^2+7 x+4\right)}{6 x} L_m^2
  -\ln (r) \Biggl[3 (x+1) \ln ^2(x)
\nn\\
&\quad  -\frac{x-1}{3 x (r x-1) \left(r x^2-1\right)^3} \Bigl(12 r^4 x^7-r^3 x^7-16 r^3 x^6-2 r^2 x^6-34 r^3 x^5-3 r^2 x^5
\nn\\
&\qquad +66 r^2 x^4+16 r x^4+20 r^2 x^3-35 r
    x^3-32 r x^2-6 x^2-6 r x+15 x+6\Bigr)
\nn\\
&\quad  -(4 r x-x-3) \ln (1-r x)
    +\frac{4 \left(x^3+3 r x^2-3 x-1\right)}{3 x} \ln \bigl(1-r x^2\bigr)
\nn\\
&\quad  -\frac{8 x^3+12 r x^2+3 x^2-15 x-8}{3 x} \ln (x)\Biggr]
  +\frac{1}{54 x \left(r x^2-1\right)^2} \Bigl[400 r^2 x^7-279 r^2 x^6
\nn\\
&\quad    +216 r^2 \zeta_3 x^6-9 r^2 x^5-656 r x^5+216 r^2 \zeta_3 x^5
    -112  r^2 x^4+126 r x^4-432 r \zeta_3 x^4
\nn\\
&\quad    +450 r x^3-432 r \zeta_3 x^3
    +400 x^3+80 r x^2+216 \zeta_3 x^2-279 x^2+216 \zeta_3 x-9 x-112 \Bigr]
\nn\\
&   -4 (x+1) \Li_3(x)  -4(x+1) \Li_3(r x)  +4 (x+1) \Li_3\bigl(r x^2\bigr)
\,,
\label{eq:phifeQf}
\end{align}

\begin{align}
   \varphi^{\two,f \neq e}_{\gamma/e} &=
  -\frac{192 (x-1)  v^2 + 40 \left(5 x^2-4 x+4\right)}{45 x}\ln (2v)
    +\frac{4 \left(5 x^2-4 x+4\right)}{9 x} \Bigl[ \ln \bigl(x^2\bigr)  +L_m \Bigr]
\nn\\
& +\frac{8 \sqrt{v^2+1} }{45 v^3 x} \Bigl[24 (x-1) v^4+\left(25 x^2-32 x+32\right)
    v^2-5 x^2+4 x-4\Bigr]\ln \Bigl(v+\sqrt{v^2+1}\Bigr)
\nn\\
&   +\frac{4 \left(2 v^4+v^2-1\right) \left(x^2-2 x+2\right)}{3 v^3 \sqrt{v^2+1} x}
\Biggl[
   \ln^2\Bigl(v+\sqrt{v^2+1}\Bigr)
   -  \ln \bigl(4v^2\bigr) \ln \bigl(v+\sqrt{v^2+1}\bigr)
\nn\\
&\quad   +  \Li_2\Biggl(\frac{\sqrt{v^2+1}-v}{v+\sqrt{v^2+1}}\Biggr) \Biggr]
-\frac{1}{3} \biggl[ x-2+\frac{2}{x}\biggr] \ln^2(1-x)
-\frac{4 \left(x^2+x-1\right)}{9 x} \ln (1-x)
\nn\\
&+\frac{2 \left(2 v^4+2 \sqrt{v^2+1} v^3+v^2-1\right) \left(x^2-2 x+2\right)}{3 v^3   \sqrt{v^2+1} x}\,
  \Li_2\biggl(\frac{2 v}{v+\sqrt{v^2+1}}\biggr)
\nn\\
&-\frac{2 \left(2 v^4-2 \sqrt{v^2+1} v^3+v^2-1\right) \left(x^2-2 x+2\right)}{3 v^3   \sqrt{v^2+1} x}\,
\Li_2\biggl(\frac{2 v}{v-\sqrt{v^2+1}}\biggr)
\nn\\
&\quad  -\frac{2}{135 v^3 x}  \Biggl[
  \bigl(65 x^2-88 x+88\bigr) v^3  -12 \bigl(5 x^2-4 x+4\bigr) v
\nn\\
&\quad
  +\frac{15 \pi ^2 \left(2 v^4+v^2-1\right) \left(x^2-2   x+2\right)}{\sqrt{v^2+1}}
    \Biggr]
  + \frac{2 \left(x^2-2 x+2\right)}{3 x} \biggl[  2 \ln \biggl( \frac{ 1-x}{x^2} \biggr) L_m
  -  L_m^2
\nn\\
&\quad+  \ln \bigl(x^4\bigr)  \ln (1-x)
  -  \ln ^2\bigl(x^2\bigr)  + \ln ^2\bigl(v^2\bigr)
  +\ln (16) \ln \bigl(v^2\bigr)   +\ln ^2(4)   \biggr]
  \,,
  \label{eq:phigameQf}
\end{align}
where we have defined
\begin{align}
  L_m \equiv \ln \biggl( \frac{m_e^2}{\mu^2} \biggr)\,,\qquad
  r \equiv \frac{m_e^2}{m_f^2} \,, \qquad
  u \equiv \frac{1-x}{2} \sqrt{\frac{r}{x}}  \,, \qquad
  v \equiv \frac{x}{2} \sqrt{\frac{r}{1-x}} \,.
\end{align}
All non-distributional terms in \eqsm{phieeNf1}{phigameQf} are understood to be multiplied by $\theta(1-x)$.
We have checked that the equal mass limit $m_f \to m_e$ of \eqs{phieeQf}{phifeQf} reproduces the corresponding one-flavor QED contributions given in \rcite{Blumlein:2011mi}.
As further consistency checks we have verified the sum rules in \eqs{chargesumrule1}{momsumrule} explicitly using the renormalized NNLO PDF results in \eqsm{f2ee}{f2game}.
In particular, we have checked that the expression for $\varphi^{\two,N_f=1}_{\gamma/e}$ we have taken from \rcite{Ablinger:2020qvo} obeys the $N_f =1$ momentum conservation sum rule in \eq{momsumruleNf1} and is thus compatible with our conventions.
For completeness we give the renormalized NLO PDFs, i.e.\  $f_{e/e}^{\one}$  and $f_{\gamma/e}^{\one}$, in \app{NLOresults}.

Finally, we briefly comment on the dependence of the NNLO electron PDFs on the masses $m_{f}$ of the additional fermion flavors ($f\neq e$).
Working in the SM (or beyond), these fermions can also be light quarks ($f = q$) with masses $m_q$, which are perturbatively not well-defined at renormalization scales $\mu \sim m_q$.
In particular, light quark flavors contribute to the flavor sums in $f^\two_{e/e}$ and $f^\two_{\gamma/e}$. The dependence of these PDFs on the light-quark masses  ($Q \gg m_q \gtrsim m_e$) signals non-perturbative corrections to $e^+ e^-$ cross sections, even if the hard scattering process is insensitive to QCD effects.
We stress that these corrections first appear at $\ord(\alpha^2)$, but are not power suppressed (although $f^\two_{\gamma/e}$ and $f^\two_{f/e}$, unlike $f^\two_{e/e}$, are regular in the limit $m_f \to 0$ with fixed $m_e$).
Whether such nonperturbative effects can be absorbed into the electromagnetic coupling via a suitable choice of renormalization scheme is an interesting question beyond the scope of this paper and should be addressed in future phenomenological studies.

\section{Conclusions}
\label{sec:conclusions}

We have computed $\ord(\alpha^2)$ corrections to the unpolarized electron PDFs in QED with an arbitrary number of massive fermion flavors directly in momentum space.
In this way we have confirmed the one-flavor QED results for the electron-in-electron PDF $f_{e/e}$ and  for the photon-in-electron PDF $f_{\gamma/e}$ obtained from a Mellin-space calculation in  \rcite{Blumlein:2011mi} and  \rcite{Ablinger:2020qvo}, respectively.
This demonstrates explicitly that the SCET definition of the PDFs, which we have used, and the traditional one in full QED are equivalent.
Our new results are the NNLO corrections to $f_{e/e}$ and $f_{\gamma/e}$ from fermion loops of a different flavor ($f\neq e$) than the electron, as well as the $\ord(\alpha^2)$ fermion-in-electron PDF $f_{f/e}$. The explicit expressions are presented in \eqsm{phieeQf}{phigameQf} and represent the main outcome of this work.

The electron PDFs are universal ingredients in the collinear factorization of cross sections of high-energy collider processes with initial-state electrons (and/or positrons).
They capture the effects of collinear QED ISR on the effective center-of-mass energy and the rapidity of the hard final state.
For a sufficiently inclusive process collinear factorization reduces the calculation of the cross section at leading order in the expansion of small fermion masses to that of the corresponding massless cross sections for the relevant partonic channels, which are then convolved with the PDFs.
We have reviewed the SCET derivation of QED factorization formulas for such processes and briefly discussed the extension to the case with resolved final-state partons (photons, light fermions) requiring fragmentation functions, the final-state analogs of the PDFs.

As fixed-order expressions our NNLO results constitute the initial conditions of the NNLL RG evolution of the electron PDFs. The NNLL running of the PDFs is governed by a DGLAP-type anomalous dimension which is determined by the known QCD splitting functions up to three loops~\cite{Vogt:2004mw,Moch:2004pa} upon conversion to QED by adjusting the color factors.
A phenomenological analysis of the NNLL resummation effects and the size of the new corrections to the PDFs when applied to W-pair~\cite{Actis:2008rb,Denner:2005fg} and top-pair threshold production~\cite{Beneke:2017rdn,Beneke:2015kwa,Hoang:2013uda} at $e^+e^-$ colliders is left to future work.

Finally, we note that with the complete set of NNLO results presented here, the electron PDFs can be directly promoted to NNLO electron beam functions. These functions encode not only the information about the longitudinal momentum of the parton participating in the hard scattering, but also about its transverse momentum ($p_\perp$), its virtuality ($t$), or both, when $m \ll p_\perp,\sqrt{t} \ll Q$.
To this end the NNLO electron PDFs have to be convolved with the QED version of the NNLO beam function matching kernels computed in \rcites{Echevarria:2016scs,Lubbert:2016rku,Gehrmann:2014yya,Gaunt:2014cfa,Gaunt:2014xga,Gaunt:2020xlc,Gaunt:2014xxa}.
When accompanied with the corresponding soft functions in dedicated factorization formulas, these beam functions might have interesting applications at future lepton (-hadron) colliders yet to be fully explored: for instance in the context of precision measurements, validation and tuning of Monte Carlo ISR models, or new physics searches via missing transverse momentum.

\begin{acknowledgments}
I am grateful to Jonathan Gaunt and Stefan Dittmaier for useful comments on the manuscript.
I wish to thank the CERN Theoretical Physics Department for hospitality while part of this work was carried out.
\end{acknowledgments}

\begin{appendix}

\section{NLO PDF results}
\label{app:NLOresults}

The one-loop computation of the bare electron PDF matrix elements yields
\begin{align}
   f_{e/e}^{\one,\bare} &= \frac{1}{\epsilon}P^\zero_{ee}(x) +  f_{e/e}^{\one}
    +\eps \, Q_e^2 \,\theta(x) \biggl[
    \delta(1-x)\biggl(\frac{3}{4} L_m^2  -2 L_m+\frac{\pi ^2}{8}+4\biggr)
 \nn\\
 &\quad    +\biggl( L_m^2+2 L_m+\frac{\pi ^2}{6}\biggr) \cL_0(1-x)
    + 4(L_m+1) \cL_1(1-x) + 4 \cL_2(1-x)
 \nn\\
 &\quad  + \theta(1-x)\, (x+1) \biggl(-\frac{1}{2}  L_m^2    -L_m  \Bigl(2 \ln (1-x)+1 \Bigr)
 \nn\\
 &\quad   - 2 \ln^2 (1-x) - 2 \ln (1-x) -\frac{\pi^2}{12}  \biggr)
    \biggr] + \ord\bigl(\eps^2 \bigr) \,,
   \\
   f_{\gamma/e}^{\one,\bare} &= \frac{1}{\epsilon}P^\zero_{\gamma e}(x) + f_{\gamma/e}^{\one}
   +\eps \, Q_e^2 \, \theta(x)\,\theta(1-x)\, \frac{x^2-2x +2}{x}  \biggl[
   \frac{L_m^2}{2}+L_m \Bigl(2 \ln (x)+1\Bigr)
 \nn\\
 &\quad +2 \ln^2(x) +2 \ln (x)+\frac{\pi ^2}{12}
   \biggr] + \ord\bigl(\eps^2 \bigr) \,,
\end{align}
with the one-loop splitting functions $P^\zero_{ie}$ as given in \app{splitting}.
The corresponding $\msb$ renormalized results are
\begin{align}
 f_{e/e}^{\one} &= Q_e^2 \, \theta(x) \biggl[
\delta(1-x) \biggl(2-\frac{3}{2}L_m \biggr)-2 (L_m+1) \cL_0(1-x)
  -4 \cL_1(1-x)
 \nn\\
 &\quad    +\theta(1-x)\, (x+1) \Bigl(L_m+2 \ln (1-x)+1 \Bigr) \biggr] \,,
 \\
 f_{\gamma/e}^{\one} &=  -Q_e^2\,\theta(x) \,\theta(1-x)\, \frac{x^2-2x +2}{x}
  \Bigl(L_m+2 \ln(x)+1 \Bigr) \,.
\end{align}

\section{Splitting functions}
\label{app:splitting}

We define the perturbative expansion of the collinear splitting functions as
\begin{align}
\label{eq:Pijexp}
P_{ij}(z,\alpha) = \sum_{n=0}^{\infty} \left(\dfrac{\alpha}{2\pi}\right)^{n+1} P_{ij}^{(n)}(z)\,.
\end{align}
The one-loop terms read
\begin{align}
P_{f_i f_j}^\zero(z) &= Q_{f_i}^2\, \theta(z)\, \delta_{ij} P_{f\!f}(z)
\,,\nn\\
P_{f_i\gamma}^\zero(z) = P_{\bar f_i \gamma}^\zero(z) &= Q_{f_i}^2\, \theta(z) P_{f\gamma}(z)
\,,\nn\\
P_{\gamma\gamma}^\zero(z) &= - \frac{2}{3} \sum_f Q_f^2\; \delta(1-z)
\,,\nn\\
P_{\gamma f_i}^\zero(z) = P_{\gamma \bar{f}_i}^\zero(z) &=Q_{f_i}^2\, \theta(z) P_{\gamma f}(z)
\,,
\end{align}
with the one-loop (LO) fermion and photon splitting functions
\begin{align} \label{eq:Pij}
P_{f\!f}(z)
&= \cL_0(1-z) \bigl(1+z^2 \bigr) + \frac{3}{2}\,\delta(1-z)
\equiv \biggl[\theta(1-z)\,\frac{1+z^2}{1-z}\biggr]_+
\,,\nn\\
P_{f\gamma}(z) &= \theta(1-z)\Bigl[(1-z)^2+ z^2\Bigr]
\,,\nn\\
P_{\gamma f}(z) &= \theta(1-z)\, \frac{1+(1-z)^2}{z}
\,.
\end{align}
%
For our NNLO calculation we need the two-loop splitting functions
\begin{align}
  P_{f_i f_j}^\one(z) &= Q_{f_i}^2 \, \theta(z) \Bigl[ \delta_{ij} P_{f_if_iV}^\one(z) + P_{f_i f_jS}^\one(z) \Bigr]
\,, \nn \\
  P_{\bar{f}_i f_j}^\one(z) = P_{f_i \bar{f}_j}^\one(z) &= Q_{f_i}^2 \, \theta(z) \Bigl[ \delta_{ij} P_{f_i \bar{f}_iV}^\one(z) + P_{f_i f_j S}^\one(z) \Bigr]
\,.
\label{eq:splitNLOferm}
\end{align}
with
\begin{align}
P_{f_i f_i V}^\one(z)
&= -\frac{10}{9} \sum_f Q_f^2\; \cL_0(1-z) (1 + z^2)
   + \delta(1-z) \biggl[ Q_{f_i}^2 \biggl(\frac{3}{8} - \frac{\pi^2}{2} + 6\zeta_3\biggr)
\nn\\
&\quad
   -\frac43 \sum_f Q_f^2 \biggl(\frac{1}{8} + \frac{\pi^2}{6}\biggr) \biggr]
   - Q_{f_i}^2 \biggl\{\frac{1 + z^2}{1 - z} \biggl[2\ln(1-z) + \frac{3}{2}\biggr]\ln (z)
   + \frac{1 + z}{2} \ln^2(z)
\nn\\
&\quad
   + \frac{3 + 7 z}{2}\ln (z) + 5 (1 - z) \biggr\}
   -\frac43 \sum_f Q_f^2 \biggl[\frac{1}{2}\,\frac{1 + z^2}{1 - z}\ln (z) + 1 - z \biggr]
\,, \nn \\
P_{f_i\bar{f}_i V}^\one(z)
  &=  Q_{f_i}^2  \biggl\{\frac{1 + z^2}{1 + z} \biggl[-4 \Li_2(-z) + \ln^2 (z)
   - 4 \ln(1+z)\ln(z) - \frac{\pi^2}{3}  \biggr]
   +2 (1 + z) \ln (z)
\nn\\
&\quad
   + 4 (1 - z) \biggr\}
\,, \nn \\
P_{f_i f_j S}^\one(z)
&= Q_{f_j}^2  \biggl[- (1 + z) \ln^2 (z) + \biggl(1 + 5 z + \frac{8}{3} z^2 \biggr) \ln (z) + \frac{20}{9z} - 2 + 6 z - \frac{56}{9} z^2 \biggr]
\,,
\end{align}
and
\begin{align}
  P_{\gamma f_i}^\one(z) = P_{\gamma \bar f_i}^\one(z) &= Q_{f_i}^2\, \theta(z)
  \Biggl(
   -\frac43 \sum_f Q_f^2 \,  \biggl\{ P_{\gamma f}(z) \biggl[\ln(1-z) + \frac{5}{3}\biggr] + z \biggr\}
\nn\\
&\quad
  - Q_{f_i}^2 \biggl\{
  P_{\gamma f}(z) \ln^2(1-z) + \bigl[3 P_{\gamma f}(z) + 2z \bigr]\ln(1-z)
\nn\\
&\quad
  + \frac{2 - z}{2} \ln^2 (z) - \frac{4 + 7 z}{2} \ln (z)
  + \frac{5 + 7 z}{2}
  \biggr\}
  \Biggr)
  \,.
  \label{eq:splitNLOgamf}
\end{align}
For simplicity we have suppressed an overall $\theta(1-z)$ multiplying the (non-distributional) terms that are regular in the limit $z\to1$.

The QCD two-loop (NLO) splitting functions were calculated in \rcites{Furmanski:1980cm, Ellis:1996nn}.
The expressions in \eqsm{splitNLOferm}{splitNLOgamf} can be obtained directly from the QCD expressions as given in \rcites{Gaunt:2014cfa,Gaunt:2014xga} by replacing
\begin{align}
  C_f \to Q_{f_i}^2\,, \quad C_A \to 0 \,, \quad T_F \to Q_{f_j}^2 \,,
  \quad \beta_0 \to -\frac43 \sum_f Q_f^2
  \,,
\end{align}
(Abelianization), cf.\ \rcites{Blumlein:2011mi,Ablinger:2020qvo}.

For the NNLO renormalization factor of the electron PDFs we also need the following non-trivial convolutions of the one-loop splitting functions in \eq{Pij}, where we again suppress a factor $\theta(1-z)$ multiplying the non-distributional terms (and do not sum over any indices):
\begin{align}
P_{f\!f}(z) \convz P_{f\!f}(z)
&= 4\cL_1(1-z) \bigl(1+z^2 \bigr)
  + 3P_{f\!f}(z) - \biggl(\frac{9}{4} + \frac{2\pi^2}{3}\biggr) \delta(1-z)
  \nn \\ & \quad
  + \bigl[- 2 P_{f\!f}(z) + 1+z \bigr] \ln (z)
  - 2(1 - z)
\,,\nn\\
P_{f\gamma}(z) \convz P_{\gamma f}(z)
&= 2(1+z)\ln (z)
  + \frac{4}{3z} + 1 - z - \frac{4}{3}z^2
\,, \nn \\
P_{\gamma f}(z) \convz P_{f\!f}(z)
&= 2 P_{\gamma f}(z) \ln(1-z)
+ (2-z) \ln (z)
+ 2 - \frac{z}{2}
\,.
\end{align}

\section{Plus distributions}
\label{app:Useful}

We define the plus distributions in the standard way as
\begin{align} \label{eq:plusdef}
\cL_n(x)
&= \biggl[ \frac{\theta(x) \ln^n (x)}{x}\biggr]_+
 = \lim_{\varepsilon \to 0} \frac{\rd}{\rd x}\biggl[ \theta(x- \varepsilon)
 \frac{\ln^{n+1} (x)}{n+1} \biggr]
\,.
\end{align}
As explained in the text, we make use of the distributional identity
\begin{align} \label{eq:plus_exp}
\frac{\theta(x)}{x^{1-\eps}} = \frac{1}{\eps}\, \delta(x) + \sum_{n = 0}^\infty \frac{\eps^n}{n!}\, \cL_n(x)
= \frac{1}{\eps}\, \delta(x) + \cL_0(x) + \eps \cL_1(x) + \ord(\eps^2)
\,.
\end{align}

\end{appendix}

\bibliography{./QEDPDFs}
\bibliographystyle{JHEP}

\end{document}